\documentclass[bibyear]{aa} 

\usepackage{color}

\usepackage{graphicx}
\usepackage{lscape}
\usepackage{txfonts}
\usepackage{natbib}
\setcitestyle{round,aysep={},yysep={;}}

\begin{document}

   \title{Gaia-DR2 extended kinematical maps}
   \subtitle{Part II: Dynamics in the Galactic disk explaining radial and vertical velocities}


\author{M. L\'opez-Corredoira\inst{1,2}, F. Garz\'on\inst{1,2}, H.-F. Wang\inst{3,4,5}, F. Sylos Labini\inst{6,7,8}, 
R. Nagy\inst{9}, $\breve{{\rm Z}}$. Chrob\'akov\'a\inst{1,2},  J. Chang\inst{10,11}, B. Villarroel\inst{12,1}}

\institute{$^1$ Instituto de Astrof\'\i sica de Canarias, 
E-38205 La Laguna, Tenerife, Spain\\
$^2$ Departamento de Astrof\'\i sica, Universidad de La Laguna,
E-38206 La Laguna, Tenerife, Spain\\
$^3$ South$-$Western Institute for Astronomy Research, Yunnan University, Kunming, 650500, 
P.\,R.\,China\\
$^4$ Department of Astronomy, China West Normal University, Nanchong 637009, China\\
$^5$ LAMOST fellow\\
$^6$ Museo Storico della Fisica
          e Centro Studi e Ricerche Enrico Fermi, 
          Compendio del Viminale, 00184 Rome, Italy\\
$^7$ Istituto dei Sistemi Complessi, Consiglio Nazionale
          delle Ricerche,  00185 Roma, Italia \\ 
$^8$ Istituto Nazionale Fisica Nucleare, Unit\`a Roma 1, Dipartimento di
          Fisica, Universit\'a di Roma ``Sapienza'', 00185 Roma, Italia\\
$^9$ Faculty of Mathematics, Physics, and Informatics, Comenius University, Mlynska dolina, 842 48 Bratislava, Slovakia\\
$^{10}$ Key Laboratory of Optical Astronomy, National Astronomical Observatories, Chinese Academy of
 Sciences, Beijing 100012, China\\
$^{11}$ Purple Mountain Observatory, the Partner Group of MPI f\"ur Astronomie, 2 West Beijing Road, Nanjing 210008, China\\
$^{12}$ Nordic Institute of Theoretical Physics, Roslagstullsbacken 23, 106 91 Stockholm, Sweden
}

   \date{Received xxxx; accepted xxxx}

  
  \abstract
  {In our Paper I, by using statistical deconvolution methods, extended kinematics maps of Gaia-DR2 data have been produced in a range of heliocentric distances that are a factor of two to three larger than those analyzed previously
by the Gaia Collaboration with the same data. It added the range of Galactocentric distances between 13 kpc and 20 kpc to the previous maps.}
   {Here, we investigate the dynamical effects produced by different mechanisms that can explain the radial and vertical components of these extended kinematic maps, including a decomposition of bending and breathing of the vertical components.
This paper as a whole tries to be a compendium of different dynamical mechanisms whose predictions can be compared to the kinematic maps.}
{Using analytical methods or simulations, we are able to predict the main dynamical factors 
and compare them to the predictions of the extended kinematic maps of Gaia-DR2.}
{The gravitational influence of Galactic components that are different from the disk, such as the long bar or bulge, the spiral arms, or a tidal interaction with Sagittarius dwarf galaxy, may explain some features of the velocity maps, especially in the inner parts of the disk. However, they are not sufficient in explaining the most conspicuous gradients in the outer disk. Vertical motions might be dominated by external perturbations or mergers, although a minor component may be due to a warp whose amplitude evolves with time.
Here, we show with two different methods, which analyze the dispersion of velocities, that the mass distribution of the disk is flared.
Despite these partial explanations, the main observed features can only be explained in terms of out-of-equilibrium models, which are either due to external perturbers or to the fact that the disk has  not had time to reach equilibrium since its formation.}
{}
  

\titlerunning{Gaia-DR2 disk dynamics}
\authorrunning{L\'opez-Corredoira et al.}

   \maketitle
%

\section{Introduction}

The connection between kinematics and dynamics has been intensively studied during the last decades of research
about the Milky Way as a Galaxy \citep{Bin08}. Different sources of spectroscopic data have been used to obtain information
about the forces that dominate the Galactic motions; for instance, with Sloan Digital Sky Survey (SDSS) \citep{moni-bidin}, Apache Point Observatory Galactic Evolution Experiment (APOGEE) \citep{Bov14}, and Radial Velocity Experiment (RAVE) \citep{Bin14} surveys. However, the important moment pertaining to the major exploitation of six-dimensional (6D) phase space (3D spatial+3D velocity) maps occurred thanks to the kinematic maps of Gaia data \citep{Gai18}(hereafter GC18); the disk being the component with better analysis prospects due to the low distance and extinction of the sources around the Sun and toward the anticenter.
The analysis by GC18 of Gaia DR2 has provided kinematical maps for Galactocentric distances
of $R<13$ kpc. \citet{Car19}, assuming priors about the stellar distribution, slightly extended the mapping out to $R=14-16$ kpc for these kinematic maps. By using the Large Sky Area Multi-Object Fibre Spectroscopic Telescope (LAMOST) or the LAMOST+Gaia surveys, other three-dimensional (3D) kinematical maps covering a similar range of Galactocentric distances were obtained  \citep{Wang2018a,wang2019a,wang2019b}.

In \citet{Lop19}(hereafter Paper I), the kinematics maps of Gaia-DR2 data were extended in a range of heliocentric distances by a factor of two to three larger with respect to GC18, out to $R=20$ kpc, by applying a statistical deconvolution  of the parallax errors based on the Lucy's inversion method of the Fredholm integral equations of the first kind. This extension to farther distances is interesting for the kinematical studies of the disk because many relevant features, aside from an axisymmetric disk in equilibrium, occur at $R>13$ kpc. The warp, flare,
and most significant fluctuations with respect to zero radial or vertical velocities, take place where the density of the disk is lower,
so it is worth extending the analysis beyond 13 kpc from the Galactic center.
The newly extended maps provide substantial, new  and corroborated information about the disk kinematics pertaining to the following: significant 
departures from circularity in the mean orbits with radial Galactocentric velocities 
between -20 and +20 km/s and vertical velocities between -10 and +10 km/s;
variations of the azimuthal velocity with position; asymmetries between the 
northern and the southern Galactic hemispheres, especially toward the 
anticenter that includes a larger azimuthal velocity in the south; and others.
This shows us that the Milky Way (MW) is far from a simple stationary configuration in rotational 
equilibrium, but it is characterized by streaming motions in all velocity components with
conspicuous velocity gradients.

In the present paper, we investigate the dynamical effects produced by different physical mechanisms that can explain the kinematic maps derived in Paper I. The paper is organized as follows: in \S \ref{.radial} and \ref{.vertical}, we describe the main features of the radial and vertical velocities obtained in Paper I. A decomposition of bending and breathing of the vertical components (i.e., the mean and the difference of the median vertical velocities in symmetric layers with respect to the Galactic mid-plane) is given in \S 
\ref{.bend.breath}. An analysis of the azimuthal velocities and the derivation of the rotation speed will be treated in another paper (\citet{Chr19}, Paper III). 
In the following sections, we  explore whether already known morphological features of the Galactic disk can explain, at least in part, the observed trends in the radial and vertical velocities. 
In \S \ref{.bar}, we analyze the effect of the Galactic bar both in the inner and the outer disk regions.
In \S \ref{.spiral}, we explore the gravitational effects produced by the overdensities associated to the spiral arms.
In \S \ref{.mergers}, the kinematical consequences, especially in vertical velocities of minor mergers or the interaction with
satellites, is explored with the help
of some N-body simulations. The distortions of the disk and their effect over the median velocities and their dispersion due
to the warp or the flare are studied in \S \ref{.warp} and \ref{sec:flare} respectively. Finally,
an interpretation of the observed kinematical features in terms of out-of-equilibrium (\S \ref{.sylos})
seems to give a possible framework for the non-null radial and vertical velocities; in this picture, the transient nature of the outer part of the disk and of the arms is related to the presence of coherent radial velocities, with both negative (i.e., contraction) and positive (i.e., expansion) signs. Discussions and conclusions are given in the last section.
The paper as a whole tries to be a compendium of dynamical factors that can be tested through the direct observable variables: the kinematic maps.
Rather than a bibliographic review, the paper acts as more so a general reflection within the wide topic of the forces acting in the disk of our Galaxy through the application of the different hypotheses to our extended kinematic maps of Gaia-DR2.

\section{Vertical and radial asymmetrical motions in Paper I}

\subsection{Radial Galactocentric velocities}
\label{.radial}

One of the main results of Paper I is that the radial component of the velocity displays considerable gradients.
In particular, the top of Fig. 8 in Paper I shows significant radial Galactocentric velocities ($V_R$)
between -20 and +20 km/s. Previous analyses in the direction of the anticenter \citep{Lop16,Tia17,Lop19b} have
already shown these departures of mean circular orbits.  However, the analysis of Paper I  has furthermore provided the azimuthal dependence of the radial velocity component, which allows one to better compare it with some possible physical scenarios. An average ellipticity or lopsidedness may be present in the highest $R$ disk \citep{Lop16,Lop19b}.
Also, a secular expansion and contraction of the disk \citep{Lop16,Lop19b} cannot be independent 
of the azimuth, but it may affect some parts of the disk; as we see in the top left corner of Fig. 8 of Paper I, the expansion and contraction affect different parts
of the disk differently: we see expansion ($V_R>0$) for azimuths $-5^\circ\lesssim \phi \lesssim 60^\circ $ and $R>10$ kpc, whereas there is contraction ($V_R<0$) for azimuths $-50^\circ \lesssim \phi \lesssim -5^\circ $ and $R>10$ kpc. Only within $|\phi |\lesssim 25^\circ $ do we have errors that are lower than 10 km/s, which makes the detection significant. 

\begin{figure*}[htb]
\hspace{1.0cm}\includegraphics[width = 4.5in]{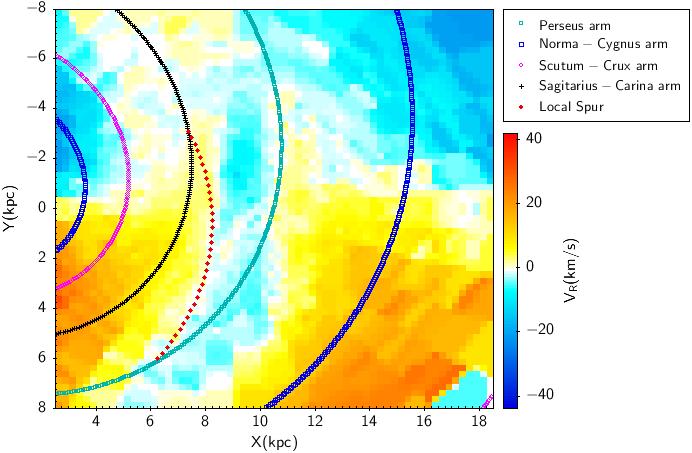}\hspace{-1.4cm}\includegraphics[width = 2.6in]{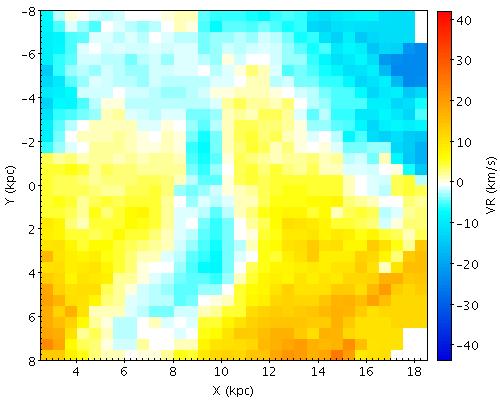}\\

\hspace{1.0cm}\includegraphics[width = 4.5in]{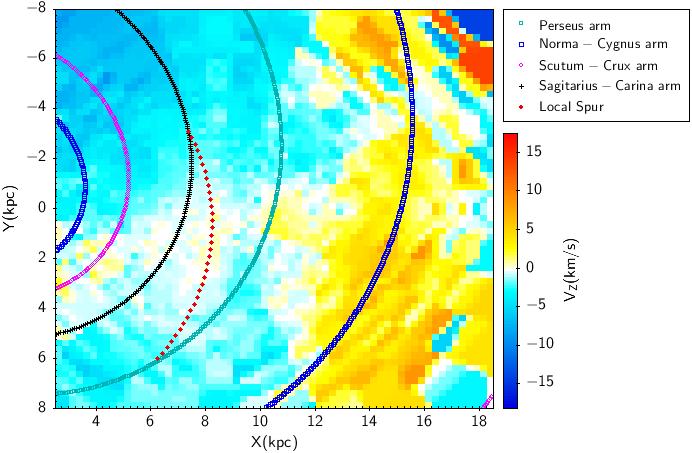}\hspace{-1.4cm}\includegraphics[width = 2.6in]{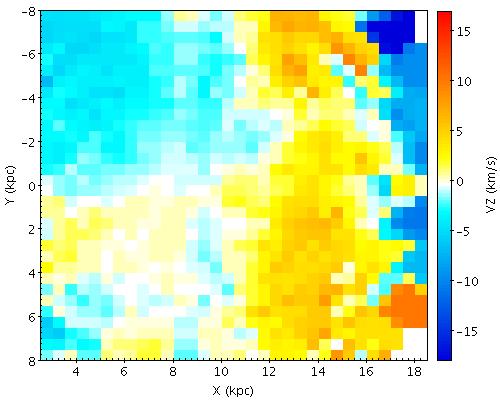}
\caption{Left: Gaia extended kinematic maps (Fig. 8 of Paper I) with the over-plotting of the representation of the spiral arms: four main spiral arms follow the model of \citet{Val17} and the local spur modeled by \citet{Hou14}: upper panel, radial velocity; bottom panel, vertical velocity.
Right: Gaia extended kinematic maps introducing a correction to the zero-point bias of parallaxes $\pi_c=\pi +0.03$ mas (Fig. 16 of Paper I).}
\label{Fig:spiral} 
\end{figure*}

We note that the effect of a zero-point systematic error in parallaxes has already been evaluated (Paper I, Sect. 4.4). Fig. 16 of Paper I shows the velocity maps with the most likely value of zero-point shift in parallaxes of -0.03 mas. and the gradients changed very slightly with respect to nonzero-point-correction. On the top of Fig. 
\ref{Fig:spiral}, we reproduce the radial velocity maps with a zoom on the region within $|Y|<8$ kpc. 
We also note that the gradients and the change of sign of $V_R$ is observed near the anticenter area (not exactly in the change of quadrant though), where $V_R$ is almost only dependent on radial heliocentric velocities and its sign would be rather insensitive to the distance determination. In any case, the propagation of errors of $\Delta r$ was taken into account, so the effect
of systematic errors in the parallaxes should not be dominant. Furthermore and as previously mentioned, in the anticenter,
the radial velocities were demonstrated to reach values up to 20 km/s in absolute value with
data that is different from Gaia \citep{Lop19b}.

Large scale motions corresponding to expansion and contraction in different parts of the outer disk are taking place, suggesting that giant processes of inflows and outflows of stars are occurring in our Galaxy. The ratio of mass exchange in these flows may be estimated with a density model of the disk together
with the information of the velocities. By assuming the extension of the flow of $\Delta \phi =40^\circ $ and by using the method introduced by \citet[\S 4.2]{Lop19b}, we get mass accretion and ejection of $\sim 5\times 10^8$ M$_\odot$/Gyr either inwards or outwards.

\subsection{Vertical Galactocentric velocities}
\label{.vertical}

The maps of vertical velocity component shown in Paper I are complex and their physical explanation has to be found in combining a variety of events. Nonaxisymmetric features have been seen previously in the kinematics of the Milky Way \citep{Fau14,Bov15}, showing a velocity distribution in the Galactic disk that is not smooth. With the use of the data from the Gaia mission (GC18, Paper I), it is now possible to extend these analyses outside the Solar vicinity  with more statistical significance. The observed velocities between -10 and +10 km/s with some correlated gradients require the existence of
some vertical forces in the Galactic dynamics.
On the bottom of Fig. \ref{Fig:spiral}, we reproduce the radial velocity maps with a zoom on the region within $|Y|<8$ kpc; the zero-point-correction of parallaxes is not important.

\section{Bending and breathing for vertical motions of the Galactic disk}

\label{.bend.breath}

\begin{figure*}
        \begin{center}
                \includegraphics[width=1.0\textwidth]{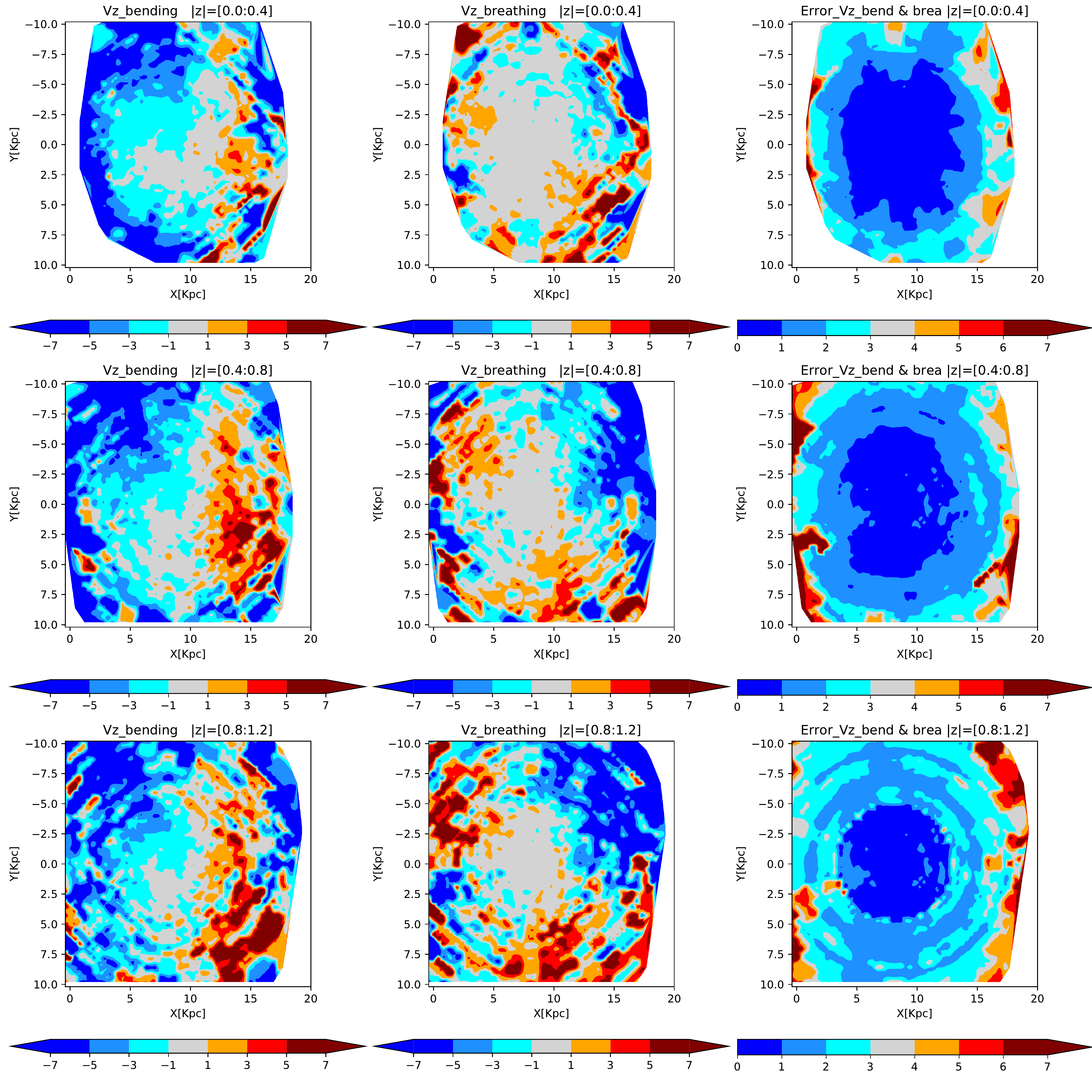}
                \caption{Bending and breathing velocities, using the same graphic layout and panel distribution as in GC18. Left column panels show the bending vertical velocity; middle column panels, the breathing vertical velocities; and right column panels, the error in the computed values for both bending and breathing modes. Units of velocities in km/s; units of scale $X$, $Y$ in kpc. Position of the Sun at $X=8.34$, $Y=0$; position of the Galactic center at $X=0$, $Y=0$. See text.}
                \label{fbb}
        \end{center}
\end{figure*}

Bending and breathing modes in the galactic disk have been discussed in \citet{1991ApJ...373..391W} as a measure of vertical oscillations of the Galactic disk.  We have used the data in Paper I to compute the bending and breathing vertical velocities at several heights over the disk, following the definitions for breathing and bending velocities in GC18 (see the paper for details):
\begin{equation}
V_{\rm bending}(X,Y;Z)=\frac{1}{2}[V_Z(X,Y,Z)+V_Z(X,Y,-Z)]
  \end{equation} and
\begin{equation}
V_{\rm breathing}(X,Y;Z)=\frac{1}{2}[V_Z(X,Y,Z)-V_Z(X,Y,-Z)]
,\end{equation}
while our analysis is extended much further away than the velocity analysis from the Solar vicinity.
In Fig. \ref{fbb}, we plot the results of the analysis following the same graphical layout of GC18 for comparison purposes. The reader must be aware that the definition of the X  coordinate is different in this paper than in GC18. In both works, the Galactic center is at $X=0$, while the Sun is  at $X=8.34$ kpc in this paper and in $X=-8.34$ kpc in GC18. Both  Fig. \ref{fbb} here and Fig. C6 in GC18 are centered around the Sun projection in the Galactic plane. The data have been binned in cells of $400\times400$\ pc in $XY$, which are then averaged in layers of 400 pc in $Z$ that are above and below the plane, in order to produce the final bending and breathing distribution. The right column panels in Fig. \ref{fbb} shows the root-mean-square (r.m.s.) errors in the velocities. The bending velocity is negative  in the inner disk, reaching values in excess of -5 km s$^{-1}$, while it is increasingly positive in the outer disk and the height of the layers are above and below the plane, with peaks over 5 km s$^{-1}$ . The breathing velocity shows a smoother distribution, which is mostly negligible in the plane, predominantly positive in the second and fourth quadrants of  Fig. \ref{fbb}, and negative in the first one. The third quadrant shows a mixture of both negative and positive values. This distribution is more marked when it is far from the plane.

In recent years, several papers have been devoted to exploring vertical motions in the Galaxy. \citet{2012ApJ...750L..41W} claim, based on the analysis of 11 000 stars
in Sloan Extension for Galactic Understanding and Exploration (SDSS-SEGUE), that the motion of stars resemble that of a breathing mode perturbation with a velocity gradient of $\sim  3\ - 5$\ km s$^{- 1}$\ kpc$^{-1}$. That would result in a negative bending velocity distribution, which is more negative with the distance off the plane and a positive breathing velocity distribution.\ This was more marked before with the distance off the plane. This is not supported by the data in  Fig. \ref{fbb}. \citet{2012ApJ...750L..41W} also show an increase of the velocity dispersion with the distance off the plane, which is attributed to a progressively larger number of kinematically hot stars from the thick disk. We show in \S \ref{sec:flare} that the flaring of the Galactic disk can explain this, at least to a certain level.

More recently, \citet{2018MNRAS.475.2679C} computed 3D velocities of a large sample of Galactic sources using radial velocities from RAVE, astrometric solution from the Tycho-Gaia Astrometric Solution (TGAS) and proper motions from several catalogs. They find differences in the velocity distributions derived from each of the proper motion data bases, which display combinations of bending and breathing modes. Our results seem to be aligned with their claim (see also \citet{Car19,Ben19}) of having detected a breathing mode inside the solar circle, which would correspond to positive breathing velocity in the second and third quadrant of Fig. \ref{fbb}, and a bending mode outside it, which would yield to a positive bending velocity in the first and  fourth quadrants. However, in agreement with \citet{Ben19}, we see clear asymmetry in the vertical velocity above and below the plane as the bending and breathing velocities do not follow antisymmetric patterns. The interior of the breathing mode to the solar circle is  more marked in the negative values of the bending velocity distribution than in the corresponding positive values of the breathing velocity in the same area. This difference is more noticeable closer to the plane. Additionally, the same thing seems to happen for the bending mode beyond the Sun; the bending velocity distribution shows a more clear positive average value than the breathing velocity distribution does with its positive value. However, due the complex nature of the observed vertical velocity distribution, it is difficult to make a reasonable description by only using a single oscillation mode. As noted in GC18, a superposition of different modes is likely to be the true distribution.

\section{The effect of the Galactic bar on the radial velocities of stars in the Galactic disk}
\label{.bar}

In recent years, several papers have investigated the effect of the central region of the Milky Way Galaxy on the radial mixing of the Galactic disk and radial migration of stars.
In this section, we intend to compare the observed features provided by kinematical maps of Paper I with results of published numerical simulations. \citet{Hal18} present a complex N-body simulation of the Galactic disk by focusing on the radial migration of stars. The model considers the Galactic bar as the strongest nonaxisymmetrical perturbation, whose angular speed decreases due to the transfer of angular momentum from the disk to the dark matter halo. The consequence of the slowing Galactic bar is an outwards-shifting of the corotation radius, thus it implies that a wider range of the Galactocentric distances is affected by the corotating resonance. The thin disk stars significantly migrate outwards 
near the corotating radius 
at approximately 10 kpc. In the case of the thick disk, the average initial radius of stars, which is influenced by the corotation resonance, is slightly shifted to lower Galactocentric distances \citep{Hal18}.
This result is in accordance with the top panel of Fig. 2 in Paper I. However, the complex behavior of radial velocities that are plotted in the top panel of Figs. 8 and 13 of Paper I cannot simply be explained by the simulations by \citet{Hal18}. One can also compare the dependence of the dispersion of the radial velocities on the Galactocentric distance \citep[Fig. A2]{Hal18} with the observational data. We can say that the closest fit to the observed radial velocity dispersion  represents the thick disk model by \citet{Hal18}, but there is no evidence that the model that was used for the Galactic bar could generate observed dispersion of radial velocities.

\begin{figure}
        \centering
        \includegraphics[width = \columnwidth]{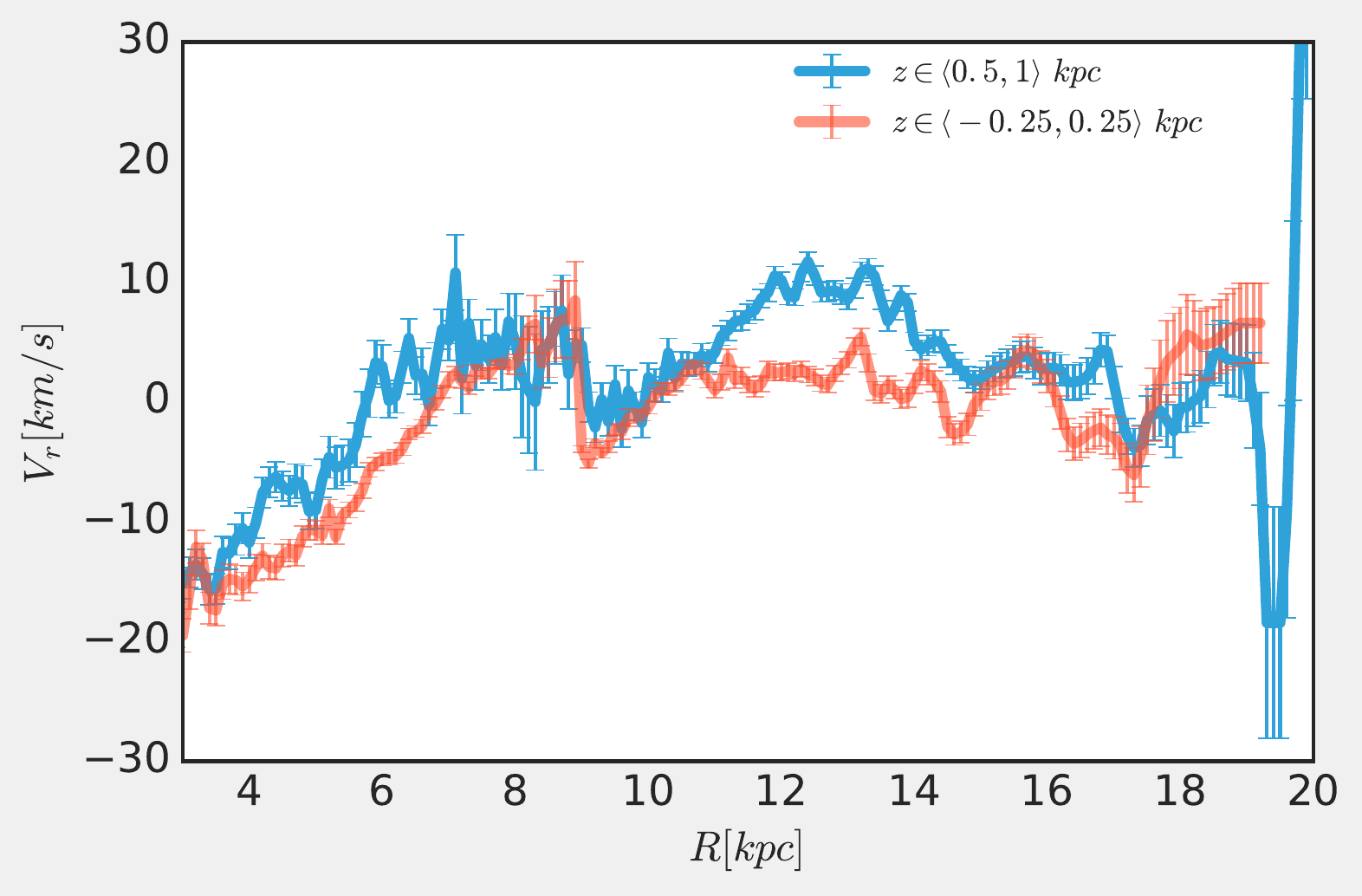}
        \vspace*{2pt}%
        \caption{Mean value of radial velocity over various $z$ intervals as a function of the Galactocentric distance along the center-Sun-anticenter direction using data by Paper I.}
        \label{bar1}
\end{figure}

\begin{figure}
        \centering
        \includegraphics[width = \columnwidth]{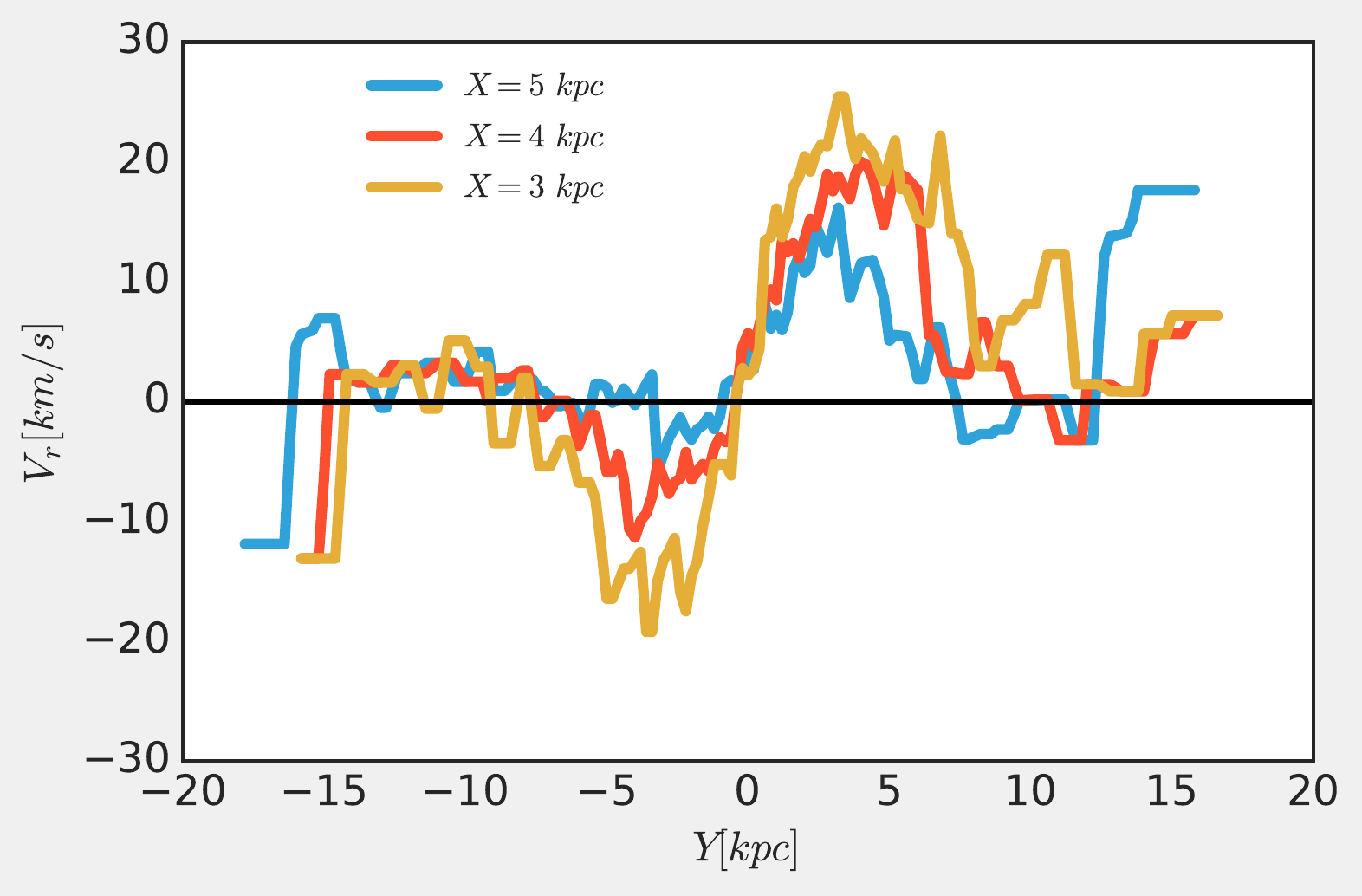}
        \vspace*{2pt}%
        \caption{Mean value of radial velocity as a function of $Y$ for various values of $X$ by using data from Paper I.}
        \label{bar4}
\end{figure}

\citet{Mon14} tested different structural parameters of the Galactic bar, but they all had a constant pattern speed of $\Omega_{b}=50$ km s$^{-1}$ kpc$^{-1}$. They consider a default bar to be a long bar and a less massive bar. The simulation covers the region around the Sun (approximately from 6 kpc to 9 kpc); the dependence of the radial velocity on the Galactocentric distance for different $Z$ in the center-Sun-anticenter direction and for various bar models is plotted \citep[Figs. 4-8]{Mon14}. 
All the results of \citet{Mon14} show the same trend, which is a decrease in the radial velocities from positive to negative values with increasing Galactocentric distances in the vicinity of the Sun. However, the behavior of the radial velocities observed in Paper I, which are plotted in Fig. \ref{bar1}, is more complex. By comparing the red line of Fig. \ref{bar1} (-0.25 kpc$<z<$+0.25 kpc) with the results of \citet{Mon14} in the considered range of Galactocentric distances, we see that the simulations cannot explain the observed features. Only for the specific range of 0.5 kpc$<z<$+1.0 kpc, the results of \citet{Mon14} are similar to the observations (refer to the blue line in Fig. \ref{bar1} with the rightmost panel of Fig. 7 in \citet{Mon14}).
\citet{Car19} also investigate the kinematics of the Galaxy by using the second data release of the Gaia mission (GC18). In the $X-Y$ plane, they observed an asymmetry in the radial velocities; a positive sign of the radial velocity in the first quadrant and a negative sign in the quadrant for $X\lesssim 5$ kpc. The same feature can also be observed in Fig. 8 of Paper I and it is clearly visible in Fig. \ref{bar4} of the present paper for various values of $X$. \citet{Car19} try to explain the pattern as a result of the presence of the Galactic bar. They could not reproduce the observed azimuthal gradient when just considering internal effects (e.g., the Galactic bar and the spiral arms). 

\begin{figure}
        \centering
        \includegraphics[width = \columnwidth]{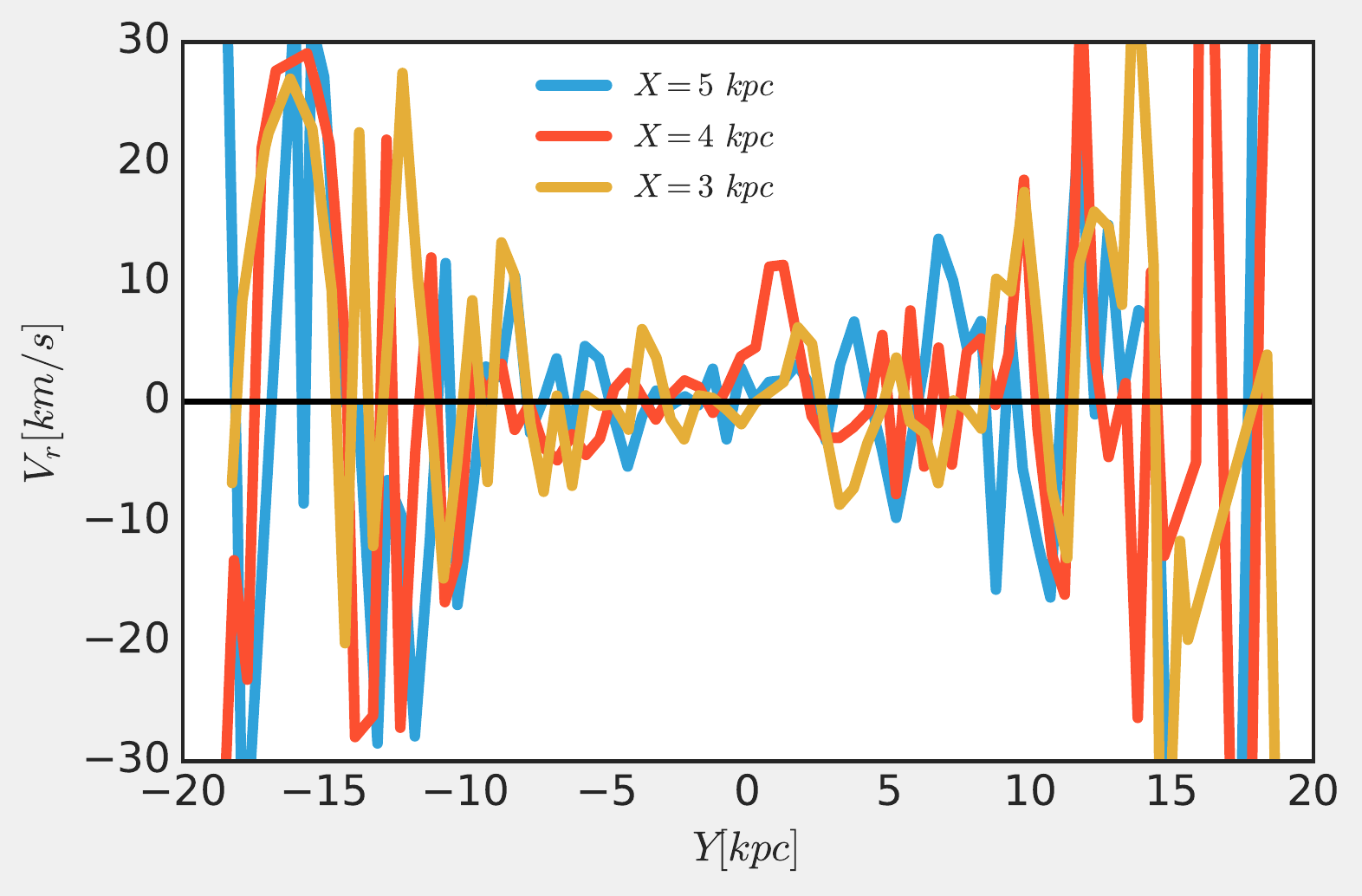}
        \vspace*{2pt}%
        \caption{Mean value of radial velocity as a function of $Y$ for various 
values of $X$ by using results of integration of $10^5$ test particles in the rotating gravitational potential of the Galactic bar considering Newtonian dynamics.}
        \label{bar2}
\end{figure}

\begin{figure}
        \centering
        \includegraphics[width = \columnwidth]{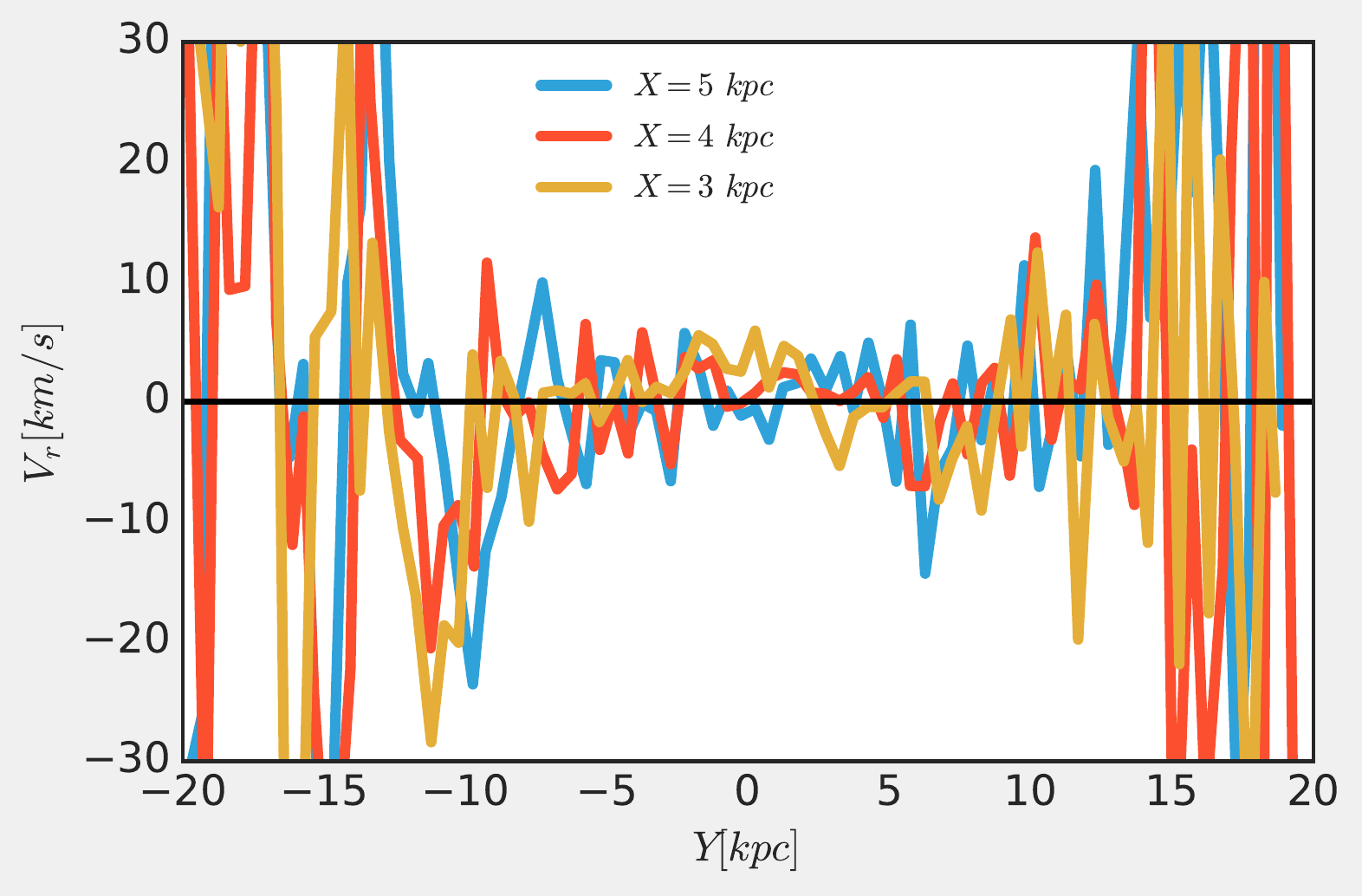}
        \vspace*{2pt}%
        \caption{Mean value of radial velocity as a function of $Y$ for various 
values of $X$ by using results of integration of $10^5$ test particles in the rotating gravitational potential of the Galactic bar considering MOND.}
        \label{bar3}
\end{figure}

We have also investigated the evolution of radial velocities by using the results of integration of $10^5$ test particles randomly distributed in the Galactic equatorial plane in the rotating gravitational potential of the Galactic bar ($\Omega_{bar}=55.5$
km~s$^{-1}$ kpc$^{-1}$, $M_{\rm bar}=9.23 \times 10^{9}M_{\odot}$) that were obtained by \citet{kac}. The simulation considered two approaches, the first one is based on the Newtonian gravitation including the Navarro-Frenk-White (NFW) dark matter halo \citep{Nav97}, the second one uses non-Newtonian gravitation (Modified Newtonian Dynamics; MOND) based on \citet{mcgaugh} without a dark matter halo that takes the following equation of motion into account \citep{kac}:
\begin{eqnarray}\label{nonNewt}
\dot{\mathbf{v}} &=&\,\frac{\mathbf{g}_{\rm bar}}{1-\mbox{exp}(-\sqrt{g_{\rm bar}/g_{+}})}~,\nonumber \\
\mathbf{g}_{\rm bar} &=& -\nabla (\Phi_{d}+\Phi_{b})~,
\end{eqnarray}
where $g_{+}=1.2\times10^{-10}\mbox{m s}^{-2}$, $\Phi_{b}$ is the gravitational potential of the Galactic bar, and $\Phi_{d}$ is the gravitational of the Galactic disk.
The results of the simulations (see Figs. \ref{bar2} and \ref{bar3}) cannot account for the observed behavior of the radial velocities seen in Fig. 8 of Paper I (a positive and negative sign of the radial velocity for $X\gtrsim 10$ kpc and azimuths $-5^\circ\lesssim \phi \lesssim 60^\circ $ and $-50^\circ \lesssim \phi \lesssim -5^\circ $, respectively). The simulations give average values of the fluctuations, which are much lower than the observed ones, and fastly varying fluctuations, which are not observed in the smooth gradient of our data.
Thus, we cannot conclude that the observed feature is caused by the effect of the Galactic bar.

\section{Spiral arms}
\label{.spiral}

When spiral arms pull stars toward them, they may generate peculiar acceleration.
This mechanism, with compression where the stars enter the spiral arm and 
expansion where they exit, was proposed as a tentative explanation for the
variations of more local radial velocities \citep{Sie11,Mon16,Gai18,Lop19b};
furthermore, some ripples and ridges in the
kinematics may be formed due to the spiral arms \citep{Fau14,Hun18,Qui18}. 
A galaxy with four spiral arms with a total mass of $\sim 3\times 10^9$ M$_\odot $ ($\sim 3$\% of the Galactic disk mass)
would give radial galactocentric velocities with an amplitude of $\sim 6$ km/s \citep{Lop19b}, 
which might be part of the explanation of the observed $V_R$.

Figure 6 of \citet{Fau14} shows that the transition when crossing a spiral arm from positive to negative velocities (or vice verse) is smooth when within short distances, and the value of $\langle V_R\rangle $ is expected to be quite stable in the inter-arm region. From this figure, we also see that the map of $V_R$ changes should follow the pattern of the spiral arms. A similar pattern is also obtained in a scenario of out-of-equilibrium formation of disk and spiral arms (see \S \ref{.sylos}), as observed in Fig. 14 of \citet{Lop19b}. 
The question is inevitably posed as to whether this is observed in our Gaia data. Fig. \ref{Fig:spiral} shows the Gaia extended kinematic maps (Fig. 8 of Paper I) with the over-plotting of the representation of the following spiral arms: four main spiral arms following the model of \citet{Val17} and the local spur modeled by \citet{Hou14}.

In the radial velocity plot, we see some possible coincidences of zero speed tracing some spiral arms, such as Scutum-Crux, local spur, or Perseus. However, some zones of transition between positive and negative velocities, such as the most
prominent one between $(X=10,Y=-5)$ and $(X=18,Y=0)$ (units in kpc) with a radial velocity gradient between these two regions of about 40 km/s, clearly cannot trace any spiral arms; indeed, this line is perpendicular to the
known spiral arms.
For the vertical velocity map, we do not see any clear association at all between spiral arms and the observed
features.
Therefore, we conjecture that spiral arms can only be a partial explanation for the radial velocities,
and some other effect should produce main features with higher amplitudes than expected from spiral arms ($\sim 6$ km/s) and with a different distribution. 
In particular, the large scale motions of expansion and contraction in different parts of the outer disk at $R\gtrsim 10$ kpc with giant processes of inflows and outflows of stars are not due to the spiral arms.

\section{Minor mergers and interaction with satellites}
\label{.mergers}

\citet{Car19} propose that a major perturbation, such as the
impact of the Sagittarius dwarf galaxy, could reproduce the observed
$V_R$ field. 
Incorporating an impact of a dwarf galaxy as an external perturbation matches the observed velocity fields. \citet{Car19} conclude that the azimuthal radial velocity gradient is strongly time-dependent and the radial velocity field will reverse after the effect of the external perturbation vanishes.
There are some indications \citep[Fig.10]{Car19} that the merger could cause another interesting feature in the radial velocity profile, such as the change of sign of the radial velocity as observed in Paper I. The comparison of the result of the simulation \citep[Fig. 10]{Car19} with the observed pattern of Paper I is not conclusive, we see some similar features but also some relevant differences.

Another perspective is given by \citet{Tia17}, who claim that it is not likely that the perturbations produced by mergers
may intensively affect the in-plane velocity. Also, a local stream cannot be the explanation for radial velocities along a wide range of $\sim 20$ kpc, 
but it could be a large-scale stream associated with the Galaxy in the Sun--Galactic center line.
We do not have evidence for such a huge structure embedded in our Galaxy, so we think this is not very likely. 
Moreover, there is a very small asymmetry of radial velocities between the northern and southern 
Galactic hemispheres (Figs. 9-12/top left of Paper I), thus we think it is very unlikely that the same stream that is independent of the Galaxy be so symmetric with respect to the plane.

Nonetheless, minor mergers may raise vertical waves \citep{Gom13}. Major mergers (1:10) along the history of the
Milky Way are excluded from chemodynamical analyses \citep{Ruc14,Ruc15}.
Also, \citet{Hai19} explore the effect of the passage of a massive satellite through the disk of a spiral galaxy and,
in particular, the induced vertical wobbles in a time varying potential.

\subsection{N-body simulations}

Here, we use a static analytic potential to model the dark matter halo of the MW, 
while particles are used to model stars in the disk and the bulge. For satellites, particles are used for both dark matter and stars. We modeled the accretion of nine satellites that infall in the first 0$-$4 \,Gyr. The dark matter halo of the MW is described as an NFW potential \citep{Nav97} with a virial mass of $M_{vir}=1\times 10^{12}\,M_{\odot}$ and a concentration parameter of $c=7$ \citep{Mac08}. The stellar part of the MW is 3\% of the virial mass \citep{Moster2013,Wang2015}, consisting of a Hernquist bulge \citep{Her90} and an exponential disk. 
The bulge contributes 20\% of the total stellar mass, and the remaining 80\% of the total stellar mass is in the disk. 
Our dynamical model is simple, we just want to investigate qualitatively how multiple minor mergers affect the asymmetry of the disk of the Milky Way and discuss the possible mechanisms of these asymmetries, so the accuracy of the parameters is not important here.
We designed three types of progenitors: H-m, M-m, and L-m, with total masses of $4\times 10^{10}\,M_{\odot}$, $1\times 10^{10}\,M_{\odot,}$ and $2.5\times 10^9\,M_{\odot}$, respectively. Using H-m, M-m, and L-m, we created nine satellites (three for each) with distinct infall scenarios by changing their initial velocities and positions. Each satellite has an NFW dark matter halo and an exponential stellar disk.
The total simulation time is 12 Gyr, then we selected all the particles with a radius from 8 to 20 \,kpc, a height from -2 to 2 \,kpc, and the stars are located in a sector of 15 degree azimuth. For more details about orbit parameters, see \citet{Yuan2018}.

\begin{figure}
  \centering
  \includegraphics[width=0.48\textwidth]{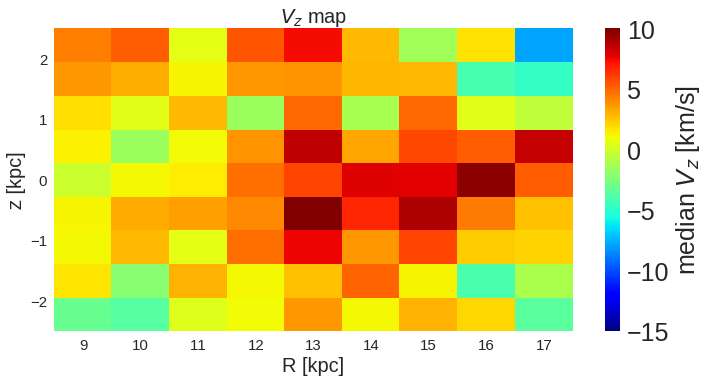}
 \caption{Edge$-$on views of the kinematics of the disk with range of 8.5$-$ 17.5 \,kpc, shown as median velocity maps of  $V_{Z}$ (in km$^{-1}$) in the N-body simulation. There are clear vertical bulk motions  or bending mode motions in the range of 10$-$17 \,kpc; especially for Z when less than 2 \,kpc.}
 \label{Nbody1}
\end{figure}

\begin{figure*}
  \centering
  \includegraphics[width=.65\textwidth]{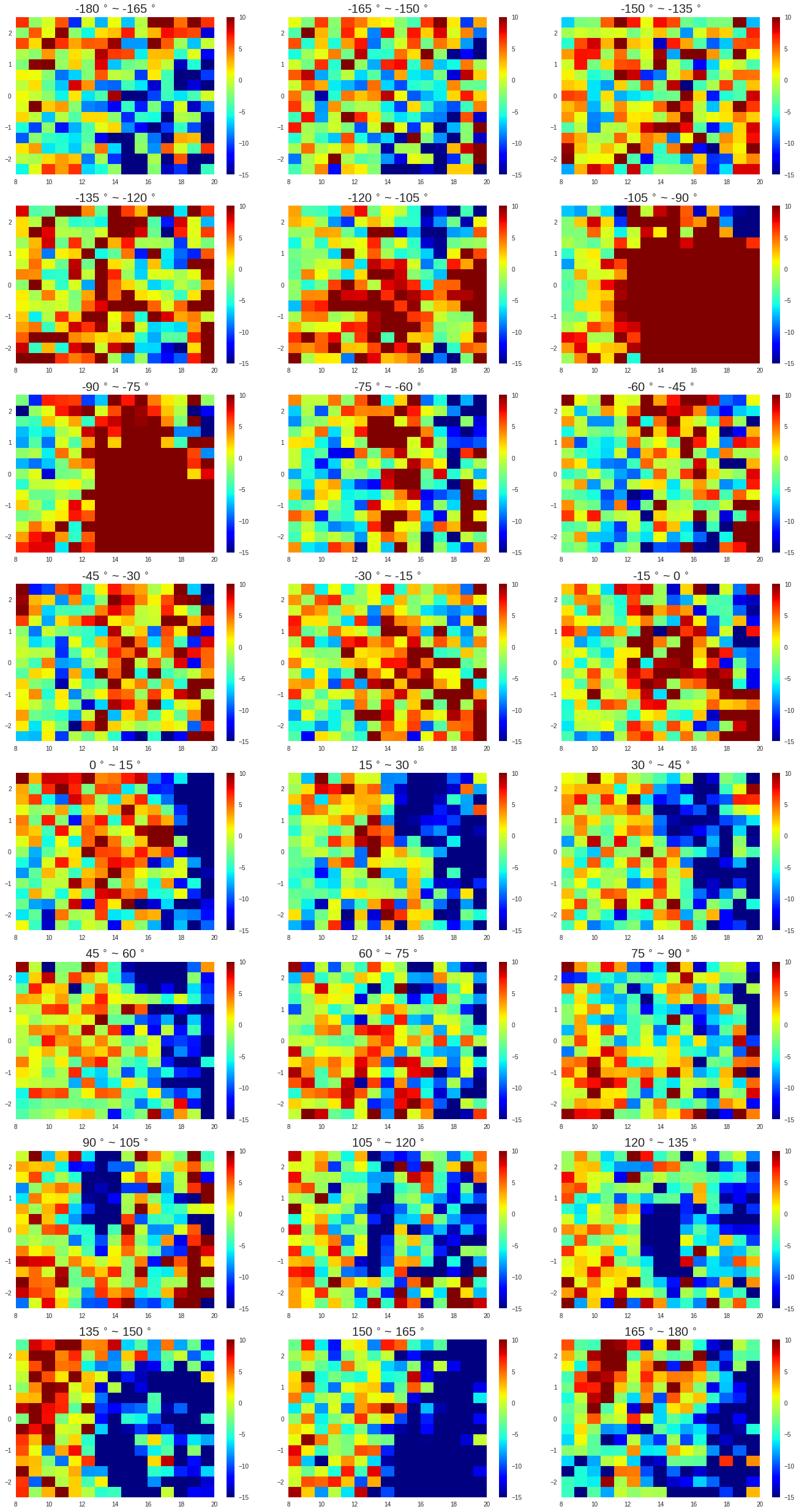}
 \caption{Edge$-$on views of the kinematics of the disk with range of $R=8$-$ 20$ \,kpc and $Z=-2.5$--2.5 kpc, shown as median velocity maps of $V_{Z}$ (in km$^{-1}$) for different azimuthal angles in the N-body simulation. Colors and labels are the same as in Fig. \ref{Nbody1}.}
 \label{Nbody_24fig}
\end{figure*}

Fig. \ref{Nbody1} shows edge$-$on views of the kinematics of the disk with a range of 8.5$-$ 17.5 \,kpc, which is shown as median velocity maps of  $V_{Z}$ (in km$^{-1}$). There are general vertical bulk motions  or bending mode motions in the range of 10$-$17 \,kpc with few negative value bins. For the distance that is less than 10 \,kpc, there are some bins with a negative or zero value;  it is matched to the general trend with Fig. 10 of Paper I by considering observational errors. The pattern of the vertical motions larger than 9 \,kpc is similar to other works using LAMOST survey \citep{Wang2018a, wang2019a, wang2019b} for the overall trend. Here we have just chosen one snapshot to investigate the effects occurring on the disk when some or many satellites interact with the galaxy. The results show that it can reconstruct recent Gaia kinematical features in some parts of the disk. Other angles are shown in Fig. \ref{Nbody_24fig}.
It is important to note, however, that there is no evidence of the existence of so many minor mergers in the Milky Way.

\section{The Galactic warp}
\label{.warp}

Warps are common phenomena in disk galaxies \citep{1991wdir.conf..181B} that have been a well known feature of the Milky Way disk for a long time, which was both observed in the gas and in the stellar components \citep[and references therein]{1993AJ....105.2138M,1998AJ....115.1483M,2000A&A...354...67D,2002A&A...394..883L,Rey09}. At present, there are several scenarios as to the origin and evolution of the Galactic warp \citep{Cas02}: interactions with the dark matter halo or nearby galaxies, infall, or others. Transient or long-lived features and some clues about its nature can be derived by using the detailed kinematics emerging from the Gaia data.
\citet{Hua18,Pog18,wang2019a} and \citet{Rom19} have already shown kinematical signatures of the Galactic warp with Gaia and other data, and here we continue such analyses within farther Galactocentric distances.

\begin{figure}
        \begin{center}
                \includegraphics[width=3.90in,height=2.94in]{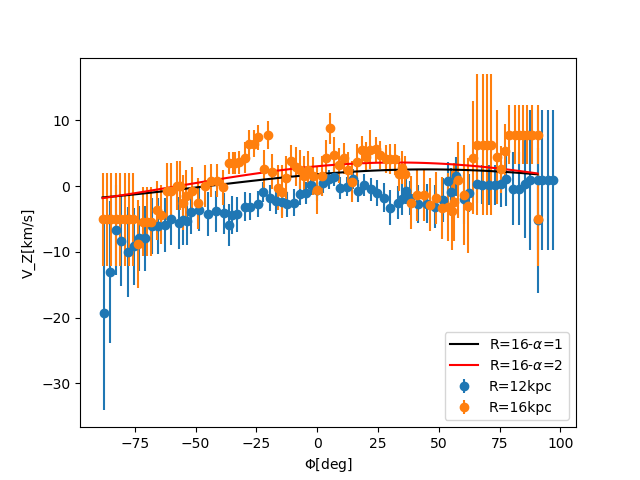}
                \caption{Median Galactocentric vertical velocities (see Paper I, Fig. 15) as a function of the Galactocentric azimuth $\phi$  ($\phi = 0$ marks the direction of the Sun). Solid line stands for the predictions of the warp model at $R=16$ kpc described in the text with $\alpha =1$ or 2.}
        \label{fwarp}
        \end{center}
\end{figure}

To this end, we used the warp model in \citet{2014A&A...572A.101L} to fit the vertical velocity ($V_{\rm z}$) distribution with the Galactic azimuth presented in Paper I (see its Fig. 15). We selected the data at Galactocentric of radii 12\ kpc and 16\ kpc.  The data were fit to the model  prediction for the vertical velocity, which
include two terms for the vertical velocity; a first one for the inclination of the orbits and the second from the temporal 
variation of the amplitude of the warp:

\begin{equation}
Max[Z_{\rm warp}]=\gamma (R-R_\odot)^\alpha
,\end{equation}
\begin{equation}
\label{eq_warp}
V_{\rm Z}(R > {R_ \odot },\phi ,z = 0) = \frac{{{{\left( {R - {R_ \odot }} \right)}^\alpha }}}{R}
\end{equation}\[\times 
\left[ {\gamma \omega _{LSR}cos(\phi  - {\phi _w}) + \dot \gamma Rsin(\phi  - {\phi _w})} \right]
,\]
with $\omega _{LSR} = 240\ {\rm km\ s}^{-1}$ being the local standard of rest rotational velocity. We tested two values for the exponent $\alpha$, 1 \citep{Rey09}, and 2, 
and we fixed the value of $\phi _w$ to $5^\circ$, known from previous works \citep[and references therein]{2014A&A...572A.101L} to avoid an unnecessary oscillation in the result. The best fit is obtained with $\chi^2= 215.64$ for $\alpha = 1$ (number of data points $N=173$) and with $\chi^2= 183.93$ for $\alpha = 2$ ($N=173$), so the quadratic model offers slightly better results. Both models together with the $V_Z$ data are plotted in  Fig. \ref{fwarp}. Only the model predictions for $R=16\ $kpc are plotted for the two values of the exponent in order to simplify the figure. As can be seen, the warp is insufficient in explaining the observed velocities, but the warp vertical velocity follows the same trend as the data. The peak to valley difference velocity for the warp model with $\alpha = 2$ in the range of azimuth between $[-75^\circ,+75^\circ]$ is 1.25\ km\ s$^{-1}$, for $R=12\ $kpc, and  4.7\ km\ s$^{-1}$ for $R=16\ $kpc. They are to be compared with 7\ km\ s$^{-1}$ and 9\ km\ s$^{-1}$ velocity difference for the data in the same azimuths and radial distance, respectively. So the warp can account roughly for between one third and one half of the observed increase in velocity.

\begin{figure}[htb]
\includegraphics[width = 3.5in]{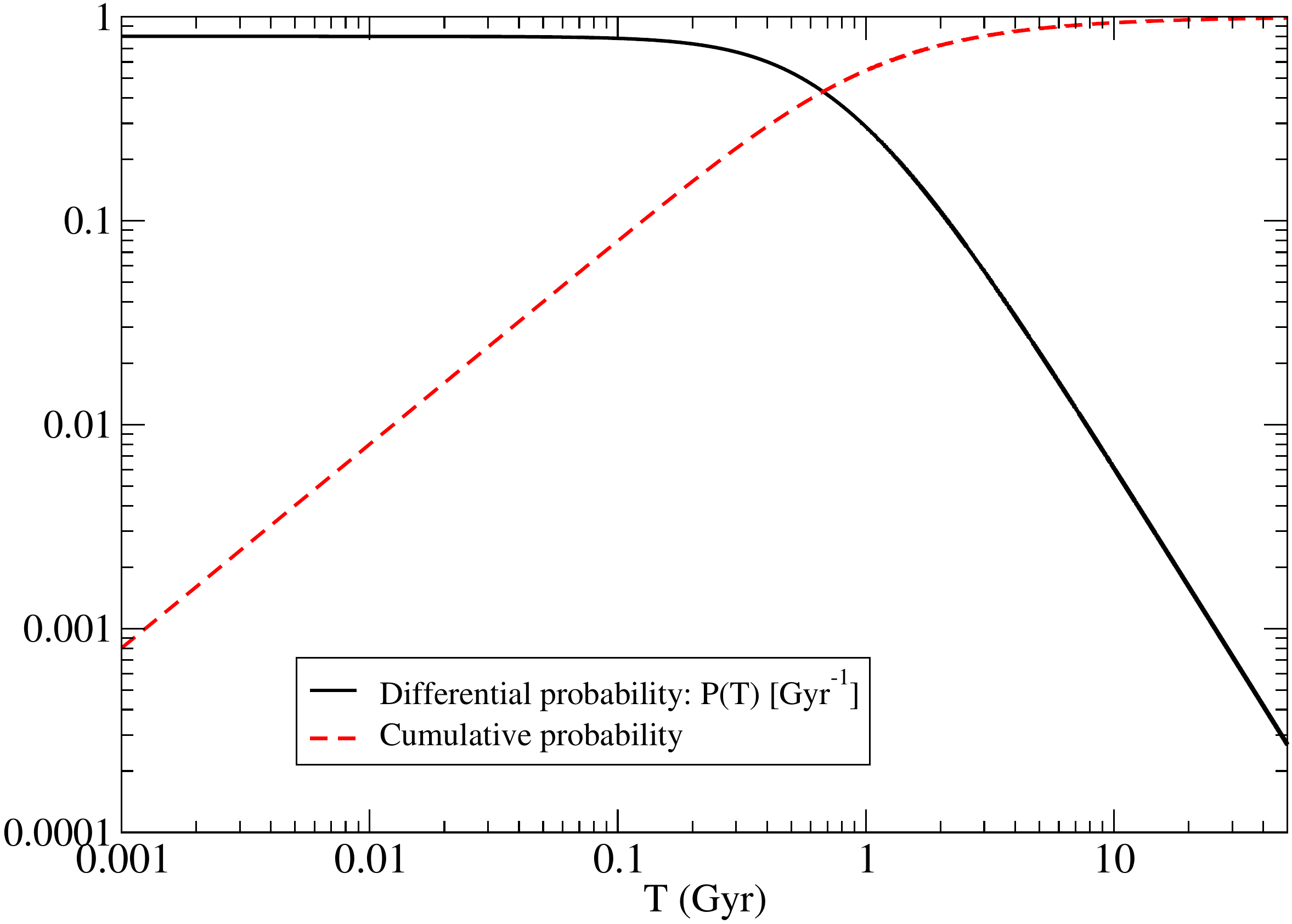}
\caption{Log-log distribution of probability of the period for the motion of $\gamma (t) = 
\gamma _{\rm max}*\sin(\omega t)$ and $\frac{\dot{\gamma}}{\gamma } = 14.6\pm 5.8$.}
\label{Fig:warpprob} 
\end{figure}

The best fit for $\alpha =1$ gives $\gamma =-0.017\pm 0.003$; 
$\frac{\dot{\gamma}}{\gamma } = 13.0\pm 5.2$ Gyr$^{-1}$. 
The best fit for $\alpha =2$ gives $\gamma =-0.0036\pm 0.0004$ kpc$^{-1}$; 
$\frac{\dot{\gamma}}{\gamma } = 7.9\pm 3.0$ Gyr$^{-1}$.  
Both values of $\frac{\dot{\gamma}}{\gamma }$ are compatible with each other, 
and they are also compatible with the values obtained by \citet{2014A&A...572A.101L} and \citet{wang2019a} .
We can derive some information about the warp kinematics. Nonzero values of $\frac{\dot{\gamma}}{\gamma }$ indicate that our warp is not stationary. If we assume a sinusoidal oscillation, $\gamma (t) = \gamma _{\rm max}\times \sin(\omega t)$, we have a period
\begin{equation}
T=\frac{2\pi }{\omega }=2\pi \left(\frac{\dot{\gamma }}{\gamma }\right)^{-1}\cot (\omega t)
\label{t}
\end{equation}
and the probability of having a period $T$ is the normalized convolution of two probability distributions \citep[Eq. 19]{2014A&A...572A.101L}:
\begin{equation}
P(T)dT=\frac{dT}{2^{1/2}\pi^{5/2}\sigma _x}\int _{-\infty}^{+\infty} dx \frac{|x|}{1+\left(\frac{Tx}{2\pi }
\right)^2}e^{-\frac{(x-x_0)^2}{2\sigma _x^2}}
,\end{equation}
where $x_0=\dot{\gamma}/\gamma $ and $\sigma _x$ is its r.m.s. Figure \ref{Fig:warpprob} shows this probability distribution for the case of $\alpha =2$. From this distribution, we can say that 
the median value is $T=0.49$ Gyr; $T$ is lower than 1.60 Gyr with 68.3\% C.L., or lower than 10.83 Gyr with 95.4\% C.L. The result
is not conclusive as to whether we have a static or variable amplitude warp.

\section{The Galactic flare}
\label{sec:flare}

It is well established that stellar kinematics is a fundamental tool to study the dynamics of disks  \citep{2011ARA&A..49..301V}. 
In this section, we use the basic assumptions behind the Jeans equation, which is that the disk is stationary and axisymmetric.\ Although, as we discuss below, there is evidence that both these assumptions do not strictly hold and they can be seen as useful working hypotheses.
Through the use of the Jeans equation, the dispersion of the vertical velocity, $\sigma_{\rm z}$, can be related to the thickness of the disk. 

\subsection{First method}

Following \citet{2010AIPC.1240..387V}, the vertical velocity dispersion for self-gravitating disks can be modeled as
\[\sigma _{\rm z}(R,z) = 
\]\begin{equation}
K\left[ h_{\rm z} \left( 2 - \exp \left(-\mid z \mid / h_{\rm z}\right)\right) \right]^{1/2} \exp\left(- (R - R_\odot)/(2h_{\rm r}) \right)
\label{eq_vzdisp}
.\end{equation}

\citet{2002A&A...394..883L} give a parametrization for a flared nonwarped Galaxy disk that uses three scale lengths: $h_{\rm r}$ as the horizontal scale length of the stellar distribution in the disk; $h_{\rm z}$ as the vertical scale of the disk that varies with $R$ in an exponential way with a scale $h_{\rm rf}$ as $ h_{\rm z} = h_{z_0}\exp\left({(R - R_\odot)} / {h_{\rm rf}}\right)$. See \citet{2002A&A...394..883L} for details. This parametrization can be combined with Eq. (\ref{eq_vzdisp}) to derive a simple model for the vertical velocity dispersion due to the flaring of the disk alone.  By grouping the radial scale lengths, $h_{\rm z}$\ and\  $h_{\rm rf}$,  in a single parameter, $H = h_{\rm rf}\ h_{\rm r} / (h_{\rm r} - h_{\rm rf})$, we end up with the equation for the velocity dispersion, $\sigma_{\rm z}$, to be fit to the data as follows:
\begin{equation}
\begin{array} {l}
\sigma _{\rm z} = K \left[ \left( 2 - \exp \left(-\mid z \mid / h_{\rm z}\right)\right) \right]^{1/2} \exp\left((R - R_\odot) / (2H) \right)
\end{array} 
\label{eqflare1}
.\end{equation}

However, the fit using Eq. (\ref{eqflare1}) produces rather poor results and we have explored several alternatives. These bad results have to be expected as Eq. (\ref{eq_vzdisp}) was derived for a pure exponential disk, without flaring that departs from exponential in off-plane regions. As noted by \citet{1988A&A...192..117V}, a pure exponential predicts too large a gradient in velocity dispersion. A modification of Eq. \ref{eqflare1}  is then proposed as,

\begin{equation}
\sigma _{\rm z}(R,z) = K^\prime  \exp \left( \mid z \mid / (2h_{\rm z})\right) \exp\left(- (R - R_\odot) / (2h_{\rm r}) \right)
\label{eqflare2}
.\end{equation}

We note that $K\  {\rm and}\  K^\prime$ in Eqs. (\ref{eqflare1}) and (\ref{eqflare2}) account for terms that do not depend on the fitted variables. The expansion series of $( 2 - \exp (-x))^{1/2}$ and $\exp( x/ 2)$ coincides with their first two terms for small values of the exponential and differs by $x^2/2-x^3/4$ considering the expansion to fourth order, so the modified function in Eq. (\ref{eqflare2}) reproduces the behavior of  Eq. (\ref{eqflare1}) in the plane, while it departs rather markedly off the plane, which is needed to reproduce the observed data.

This model in Eq. \ref{eqflare2} was then fit to the data in the anticenter region of Paper I, Fig. 9 bottom panels, from $X=8.4\ $kpc onwards. Data were averaged in bins of $0.5\ $kpc in $Z$ and $0.2\ $kpc in $X$. The model in Eq. (\ref{eqflare2}) is rather simple and cannot account for the full variability of the observed data. Thus, to avoid oscillations in the model and maintain the physical meaning of the magnitudes, the fitted parameters ($h_{\rm z_0},\ h_{\rm rf}\ {\rm and}\ H$) were restricted to a maximum value of 100\%  above of the corresponding values obtained in \citet{2002A&A...394..883L}, while $K^\prime$ was left free.

The best fit is for $\chi^2= 12768.4367$ with 328 data points. The values for the fitted parameters are: $h_{\rm z_0}=0.3 \pm 0.02 {\rm kpc;}\ h_{\rm rf}=5.72 \pm 0.34 {\rm kpc; and}\ H=3.05 \pm 0.13$\ kpc, from which a value of $h_{\rm r}=4.90$\ kpc was derived.

%

\begin{figure}
        \begin{center}
                \begin{minipage}[t]{.49\textwidth}
                        \includegraphics[width=1\textwidth]{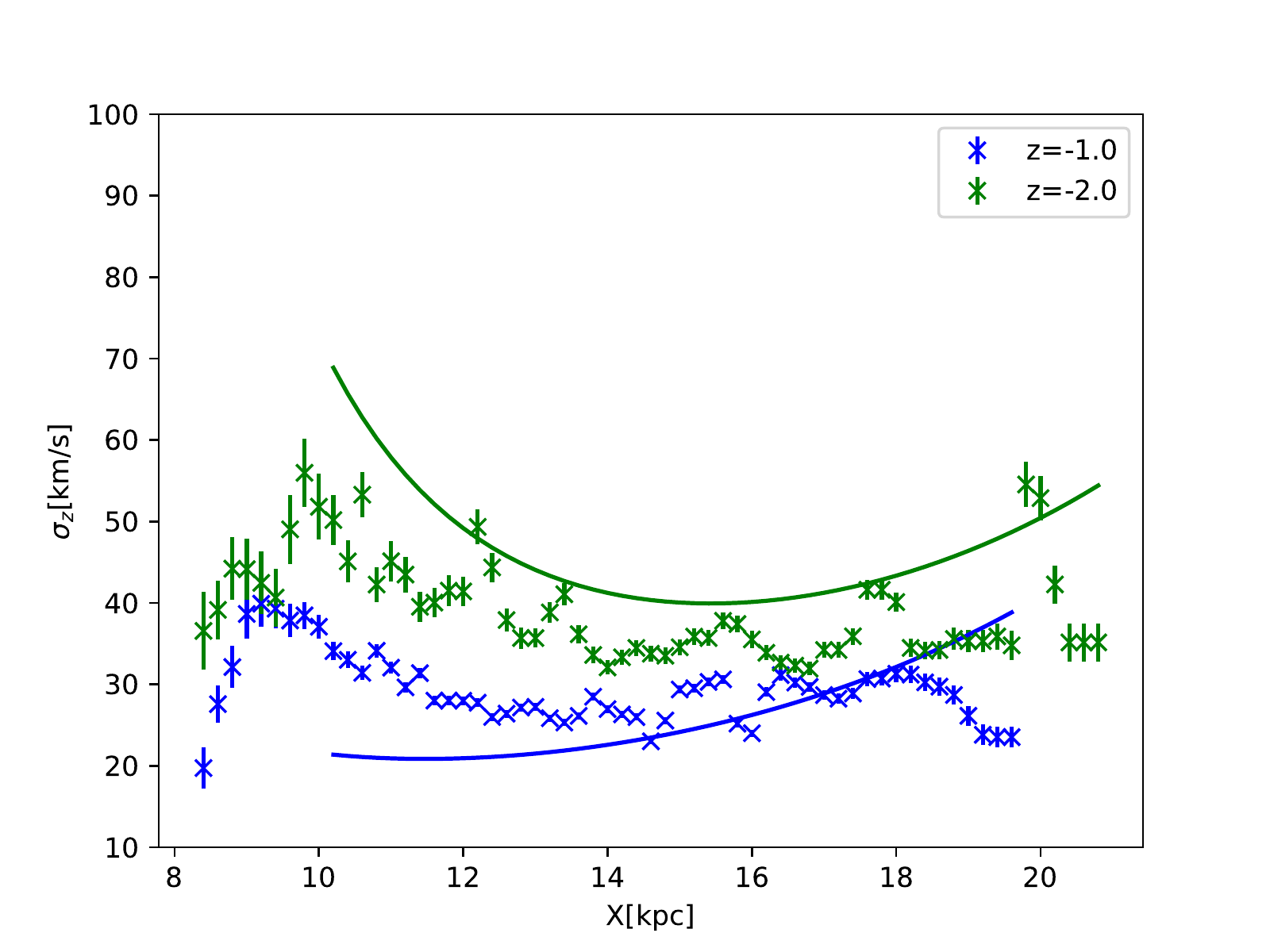}
                \end{minipage}
                \begin{minipage}[t]{.49\textwidth}
                        \includegraphics[width=1\textwidth]{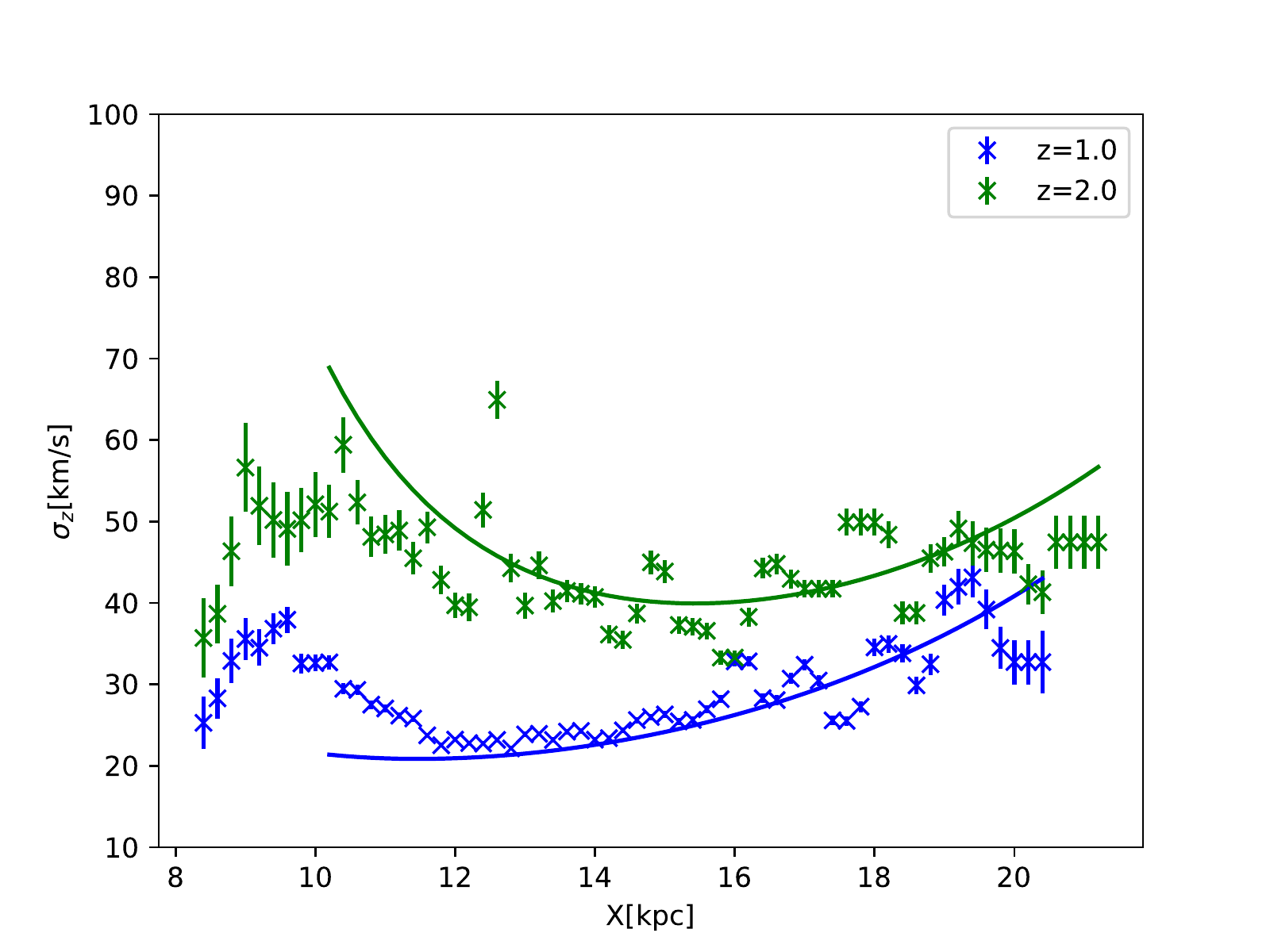}
                \end{minipage}
        \caption{Vertical velocity dispersion for different values of $Z$ (taken from Paper I, Fig. 9) with respect to the X Galactic coordinate. Data from Paper I with their error bars, are represented by crosses; the solid lines are the flare model predictions. 
Upper panel for negative $Z$ values, and lower panel for positive $Z$ values. Same color means same $Z$.}
        \label{fflare}
        \end{center}
\end{figure}

As can be seen in Fig. \ref{fflare}, the agreement is not perfect in any case, and it is better for $z >0$ than for $z < 0$, but in all cases, the trend of the dispersion velocity in the flare model is the same as in the data. Hence a fraction of the observed velocity dispersion can be attributed to the flaring of the disk.

\subsection{Fit of $h_z$ with Moni-Bidin et al. method}\label{ch6}

We can make a different use of Jeans equations to find a fitting value for the scaleheight $h_z$. We adopted the approach from \citet{moni-bidin}, who use two components of Jeans equations to express surface density from Poisson equation in cylindrical coordinates. After making a few assumptions, they obtained an expression for the surface density:

\begin{eqnarray}\label{12}
\Sigma(z)=\int_{-z}^{z} \rho \mathrm{d}z=\nonumber \\
\frac{1}{2\pi G}\left[k_1\cdot \int_{0}^{Z} \sigma_{v_R}^2 \mathrm{d}z+ k_2\cdot \int_{0}^{Z} \sigma_{v_{\phi}}^2 \mathrm{d}z
\right. \nonumber \\
\left. \cdot
 + k_3\cdot \overline{v_Rv_z}+\frac{\sigma_{v_z}^2}{h_z}-\frac{\partial\sigma_{v_z}^2}{\partial z}\right],
\end{eqnarray}
where $k_1,k_2,$ and $k_3$ are constants defined as

\begin{eqnarray}
k_1&=&\frac{3}{R_\odot \cdot h_R}-\frac{2}{h_R^2}, \nonumber \\
k_2&=&-\frac{1}{R_\odot \cdot h_R},  and \\
k_3&=&\frac{3}{h_R}-\frac{2}{R_\odot}. \nonumber
\end{eqnarray}

\begin{figure}
        \includegraphics[width=0.49\textwidth]{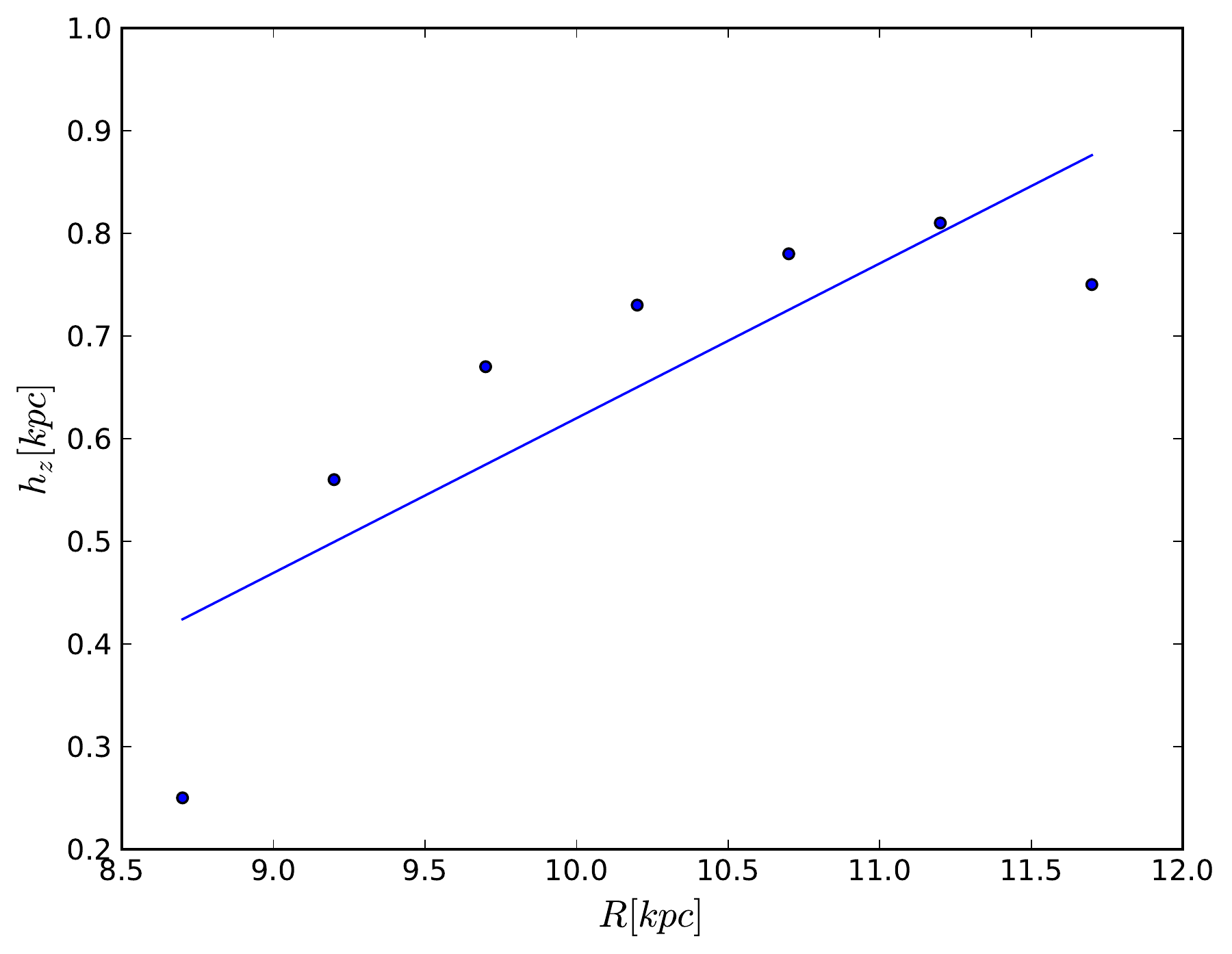}
        \caption{Fit of $h_z$ as a function of Galactocentric radius derived with Eq. (\ref{12}).}\label{o14}
\end{figure}

We calculated $\Sigma(z)$ for every $R$, choosing an initial value of $h_z=0.3$ kpc. We then proceeded to find the best value of $h_z$ for each $R$, using an iterative method as follows. We did a least squares fit of the theoretical expression $\Sigma=2\rho(R,z=0)\cdot h_z(R)\cdot[1-e^{-z/h_z(R)}]$ with every value of $h_z$ from the interval $h_z\in[0,2]$ with step $\Delta h_z=0.01$. For each value of $R$, we chose a new value of $h_z,$  which corresponds to the smallest $\chi^2$. We used this new value of $h_z$ to calculate $\Sigma$ again and we performed a new fit with theoretical the expression for $\Sigma$, which yields a new values of $h_z$ with minimal $\chi^2$ again. We repeated this procedure, until $h_z$ converged. We divided $R$ into bins with a size of $0.5$ kpc and find the best value of $h_z$ for each bin, which gives a dependence $h_z(R)$ that we show in the Fig. \ref{o14}. We fit $h_z(R)$ with the linear function, which gives values $h_z(R)=[(0.370\pm 0.093)+(0.151\pm 0.044)(R({\rm kpc})-R_\odot )]$ kpc, $R_\odot=8.34$ kpc. We only calculated $h_z$ up to $R=12.0$ kpc, as for higher values the fit was imprecise and gave untrustworthy results. We did not use a weighted fit, 
as the error of $\Sigma(z)$ is very large, which leads to an incorrect fit. \\
In deriving Eq. (\ref{12}), it was assumed that $\frac{\partial h_z}{\partial R}=0$. But, since we are interested in the effect of the flare, we needed to include terms that have not been included so far.\ This includes dependence $h_z(R)$ to Jeans equations and we recalculated the expression for $\Sigma$. Then Eq. (\ref{12}) changes as follows:

\begin{eqnarray}\label{13}
\Sigma(z)=\frac{1}{2\pi G}\left[k_1\cdot \int_{0}^{Z} \sigma_{v_R}^2 \mathrm{d}z+ k_2\cdot \int_{0}^{Z} \sigma_{v_{\phi}}^2 \mathrm{d}z + k_3\cdot \overline{v_Rv_z}\right. \nonumber \\
\left. +\frac{\sigma_{v_z}^2}{h_z}-\frac{\partial\sigma_{v_z}^2}{\partial z} +|z| \overline{v_Rv_z}\frac{\partial}{\partial R}\left(\frac{1}{h_z}\right)
\right. \nonumber \\
\left. 
+\int_{0}^{Z}|z|\frac{\sigma_{R}^2}{R}\frac{\partial}{\partial R}\left(\frac{1}{h_z}\right)+\int_{0}^{Z}|z| \sigma_{R}^2\frac{\partial^2}{\partial R^2}\left(\frac{1}{h_z}\right)\right]
.\end{eqnarray}

\begin{figure}
        \includegraphics[width=0.49\textwidth]{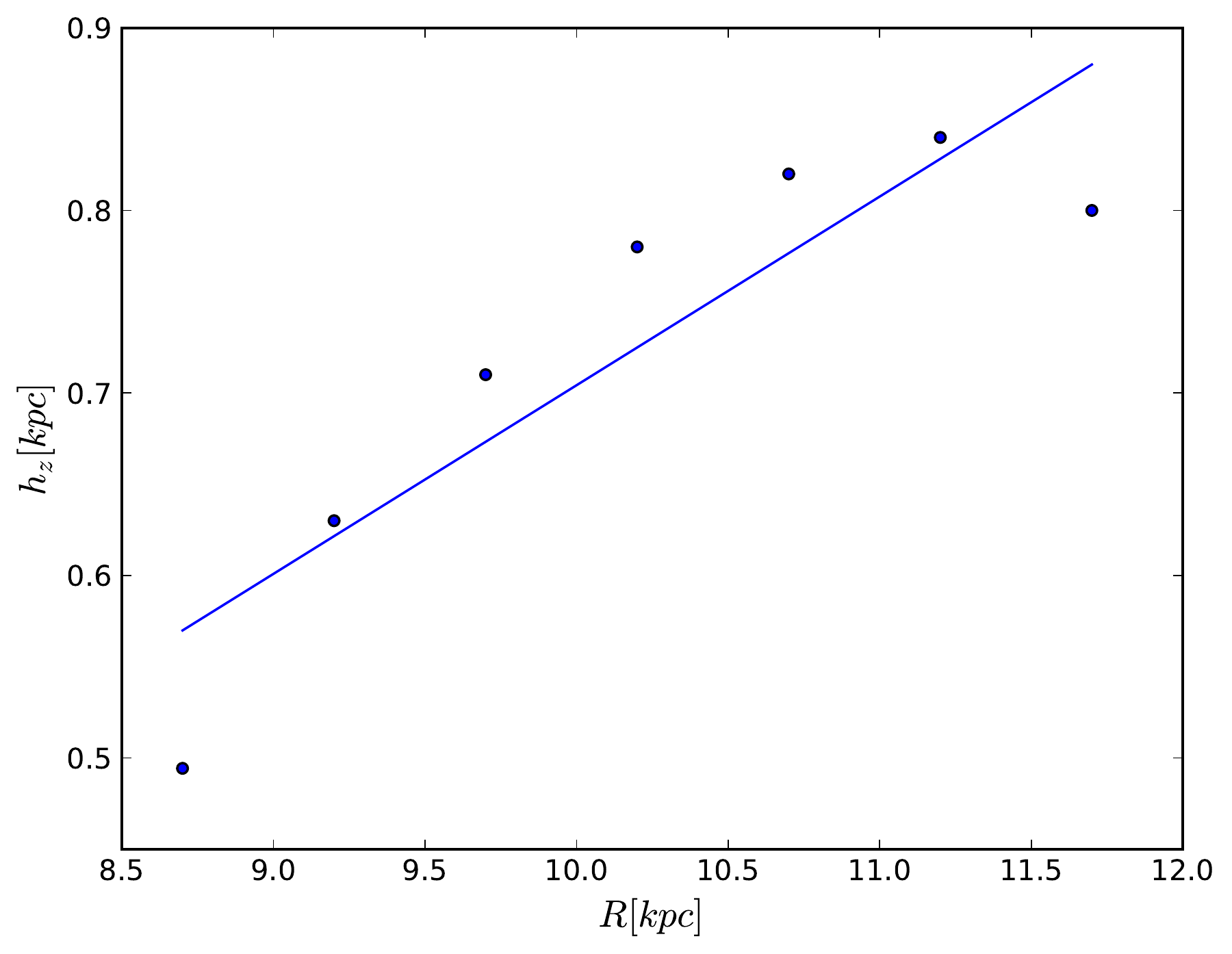}
        \caption{Same as Fig. \ref{o14}, but after correction with flare as given in Eq. (\ref{13}).}\label{o15}
\end{figure}

We repeated the iterative method for new values of $\Sigma$ and find a new dependence of $h_z(R)$. To determine the derivatives of $1/h_z$ in Eq. (\ref{13}), we used the first result for $h_z$, $h_z(R)=[0.370+0.151(R({\rm kpc})-R_\odot )]$ kpc, derived it, and plugged it in Eq. (\ref{13}). The resulting relation for $h_z(R)$ is plotted in Fig. \ref{o15}. We fit the new expression for $h_z(R)$ with the linear function 
$h_z(R)=[(0.533\pm 0.049)+(0.103\pm 0.023)(R({\rm kpc})-R_\odot )]$ kpc. \\
We see that the effect of the flare causes an increase in $h_z$ at the solar radius and decreases the slope of fit of $h_z(R)$. Moreover, it is clear that the scale height increases with distance, which is in agreement with results of other authors \citep{momany,Rey09,martin_flare,Wan18b,Wan18c}. The effect of the flare is most dominant at high Galactocentric distances $(R>15\ \mathrm{kpc})$, where our method unfortunately gives imprecise results, so we cannot compare these findings. We must bear in mind, however, that this kinematic method gives us information about the mass density (including dark matter) in all components, and not only the
stellar density in the disk.

\section{A violent origin of the Galactic disk}
\label{.sylos}

As we have discussed in the previous sections, the extended analysis
of the Gaia mission data has shown that the velocity field of the
Milky Way presents large scale gradients in all velocity components.
In particular, the radial velocity shows an azimuthal gradient on the
order of $40$ km/s in the galactic plane, ranging from $\sim - 20$ km/s
for 10 kpc $\le X \le$ 20 kpc and 0 kpc $\le Y \le$ 20 kpc to $\sim
20$ km/s for 10 kpc $\le X \le$ 20 kpc and $-20$ kpc $\le Y \le$ 0 kpc
(where $X,Y$ are the galactic coordinates) (Paper I).  The
tangential velocity also shows a gradient of $40$ km/s from the
orthogonal to parallel direction toward the anticenter.  Finally the
vertical velocity shows an azimuthal gradient on the order of $\sim$
20 km/s.  The presence of such streaming motions with conspicuous
velocity gradients implies that the Milky Way is far from a simple
stationary configuration in rotational equilibrium and that
axisymmetry is broken.
 
The dynamical origin of such morphological and velocity features
 represents an open question that has been investigated by several
 authors \citep{Antoja_etal_2018,Binney_2018}.  For
 instance, \citet{Antoja_etal_2018}, by studying several phase-space
 structures in the solar neighborhood, conclude that the
 Galactic disk is dynamically young so that modeling it as
 time-independent and axisymmetric is incorrect. In addition they have
 proposed that the disk must have been tidally perturbed between 300 Myr
 and 900 Myr ago, which is consistent with estimates of the previous
 pericentric passage of the Sagittarius dwarf galaxy.  In other words,
 the Galactic disks show features that have been generated less than
 1 Gyr ago.
 
 Here, we consider a different perspective about the origin of such
 nonstationary features;
 instead of the effect of tidal interactions
 with neighborhood satellites, we consider a model in which the
 galactic disk is isolated and has had a relatively violent origin
 from the monolithic collapse of a protogalactic cloud. 
Although a model of such a mechanism has not been yet developed in detail for the case of a realistic galactic system, the  qualitative results of such collisionless and monolithic 
 gravitational collapses, with and without gas
 dissipational gas dynamics,  are interesting because one can single out
 some unambiguous signatures of their realizations, as we are going to
 argue in what follows.

In typical cosmological scenarios (e.g., cold dark-matter-type
scenarios), structure formation proceeds hierarchically from very
small scales so that self-gravitating particles form quasi-spherical
structures, called halos, which have a small angular momentum and a
close to isotropic velocity dispersion.  The key element in the
formation of a disk galaxy is the dissipation associated with
nongravitational processes. In particular, gas can dissipate  
energy  through radiative cooling. 
The effect of such an energy decrease is that it induces the gas to collapse preferentially along its rotational axis, so that it becomes progressively flatter and thus forms a
disk. On the other hand, self gravitating particles
form through a bottom-up hierarchical aggregation process involving
a quasi-spherical configuration with
a quasi-isotropic velocity dispersion, that is, the so-called halo structures
\citep{Nav97}. 
For this reason,
dissipational effects became a necessary ingredient for the formation
of disk-like systems. In these kinds of scenarios of disk galaxy
formation, there are not rapid and large changes in the system's
physical parameters (mass, size, gravitational potential, etc.).  The
velocity field of both the gas and the self-gravitating particles,
although dominated by rotational and isotropic velocity, respectively,
is quiet, that is,  there are not large gradients and streaming
motions. 
In particular, the radial velocity is close to zero and the
different components of the systems, that is, the halo and the disk, have
reached a steady configuration \citep{Lop19b}. In this situation
only an external field, as the one induced by tidal interactions, may
generate large scale streaming motions.

When galaxy formation occurs via a top-down monolithic collapse of an
isolated over-density, the dynamical evolution is characterized by a
phase of rapid and relevant changes of the system's physical
parameters. Given the different dynamical history, the main
kinematical features of the states formed are different from those
that arise in the hierarchical scenario.

It is well known, since the pioneering work of  \citet{LyndenBell}, that 
an isolated self gravitating overdensity  in a nonstationary
configuration that rapidly changes its macroscopic properties under the
effect of the variation of its own mean field potential 
until eventually reaching a configuration that is close to a steady state.
However, the time scale for a complete relaxation from a generic
out-of-equilibrium configuration to a quasi stationary state is poorly
constrained both from a theoretical and a numerical point of view.

To explore such a dynamical mechanism,
\citet{Benhaiem+Joyce+SylosLabini_2017,Benhaiem+SylosLabini+Joyce_2019}
studied the properties of the systems formed from the pure
gravitational evolution of relatively simple initial conditions
(IC). Such systems, although extremely simplified, can give
interesting insights as to the dynamics of the monolithic collapse.  In
this way, it has been shown that when the initial conditions break
spherical symmetry and have a nonzero angular momentum, 
new and almost stationary states appear
 in which part of the mass
continues to evolve for time scales longer than the characteristic
collapse time scale $\tau \sim (\sqrt{G \rho})^{-1}$ (where $\rho$ is
the system's average mass density).  In particular, long-lived non-stationary transients formed with a rich variety of
morphological structures, such as spiral arms, bars, shells and even
rings that are qualitatively similar to those observed in spiral galaxies. A
main feature of such structures is that they are dominated by radial
motions that prevent the (fast) relaxation to an equilibrium
configuration.

When the IC break spherical symmetry, the gravitational collapse is
anisotropic and proceeds faster in the direction of the initial minor
semiaxis.
\footnote{The simplest numerical
experiments consider ellipsoidal IC where the three semiaxis are $a \ge b \ge c$.
More complex shapes must be characterized by the three eigenvalues and eigenvectors
of the inertia tensor.}
On the other hand, particles that are initially placed along the
major semiaxis take longer to arrive at the center and thus they
move for a certain time interval in a rapidly varying
gravitational field generated by the largest fraction of the mass,
which is already re-expanding. In this way, such particles gain kinetic
energy in the form of a radial motion so that, although the total
energy is conserved, the particle energy distribution largely changes.

Qualitatively the velocity field that results from these simple
systems is heterogeneous in nature and strongly scale dependent. In
general, there are three regions with different kinematic properties:
at small distances from the system center, particles form
an extended flattened region that rotates coherently, which is characterized by a 
relatively large velocity dispersion and for this reason 
the disk in rather thick. On the other hand,   
 the outermost regions
are not axisymmetric and have a radial velocity field directed
outwards, which is strongly correlated with the major axis of the system; their
energy is still negative but it is close to zero and thus their
relaxation time is very long.  The extended flattened region dominated
by circular motion arises because the collapse is more efficient along
the direction parallel to the angular momentum: it is along this
direction that the system contracts more.

When  gas dynamics is added in the 
monolithic collapse of an isolated overdensity, different behavior in the inner disk appears. Indeed,  
the gas is subjected to compression shock and radiative cooling with consequent kinetic and thermal energy dissipation so that it develops a much flatter disk, where rotational motions are coherent and the  velocity dispersion is smaller than that of 
purely self-gravitating particles. In addition, 
around such a gaseous disk long-lived, but nonstationary, spiral arms formed.
On larger scales, where the radial velocity component is significantly larger than the rotational one, the gas follows the same out-of-equilibrium spiral arms traced by 
purely self-gravitating particles  
\citep{FSL_2019}. 
Thus, even in this case, the violent dynamics characterizing 
the monolithic collapse of an isolated overdensity 
naturally give rise to a rather heterogeneous velocity field
in which motions are not maximally rotational but characterized
by large scales streams in the three velocity components with a 
predominance of radial motions in the outermost regions of
the systems. 

\begin{figure*}
\includegraphics[width = 1.42in]{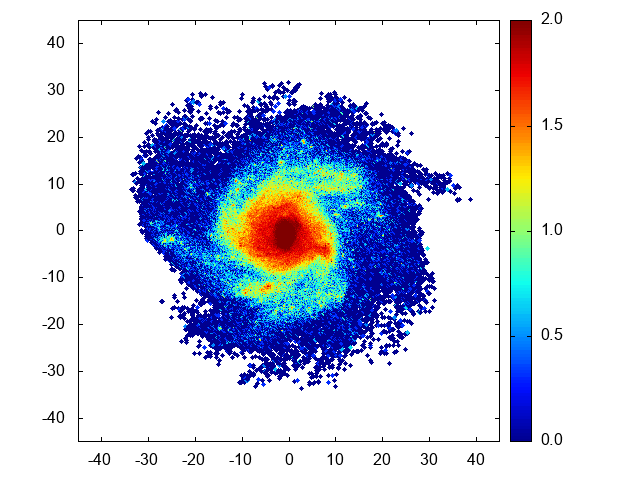}
\includegraphics[width = 1.42in]{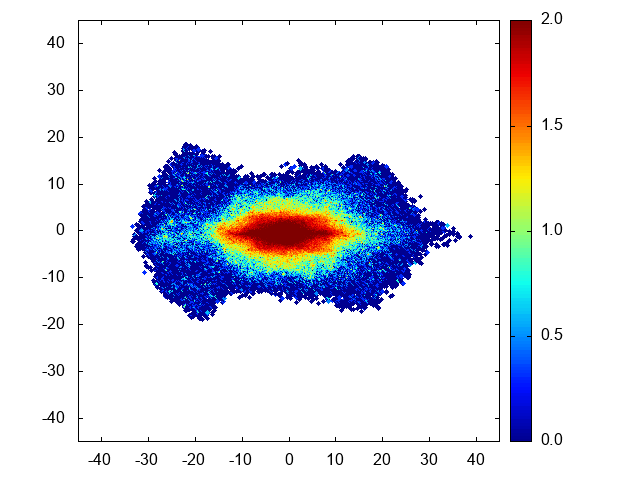}
\includegraphics[width = 1.42in]{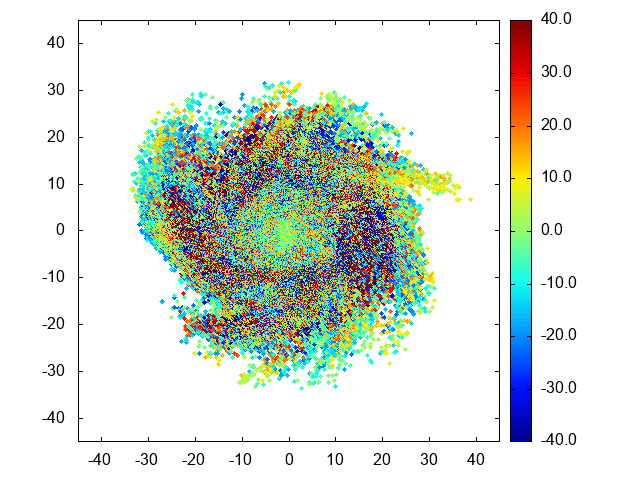}
\includegraphics[width = 1.42in]{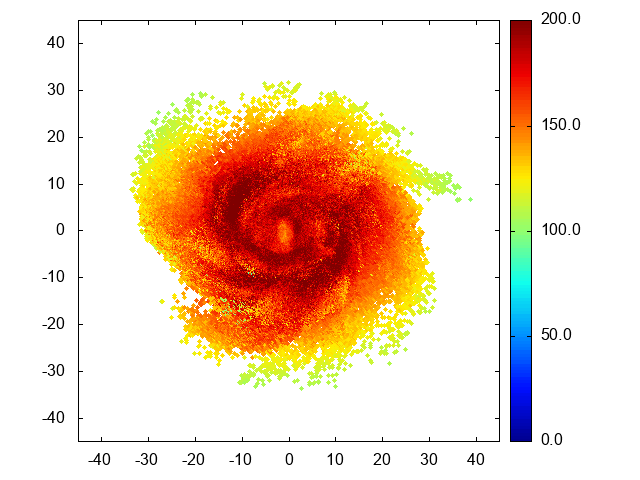}
\includegraphics[width = 1.42in]{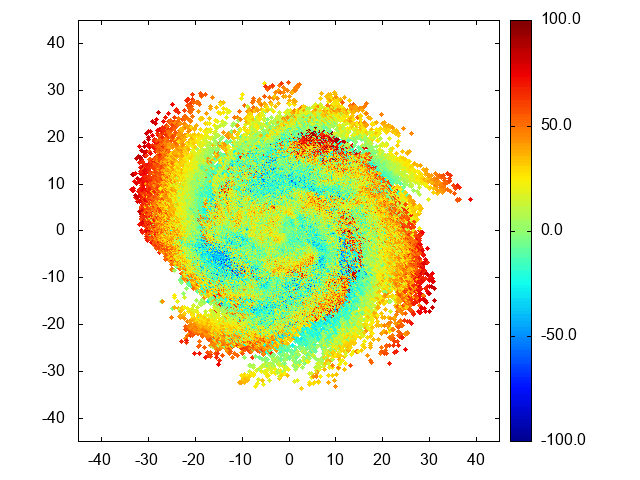}\\
\includegraphics[width = 1.42in]{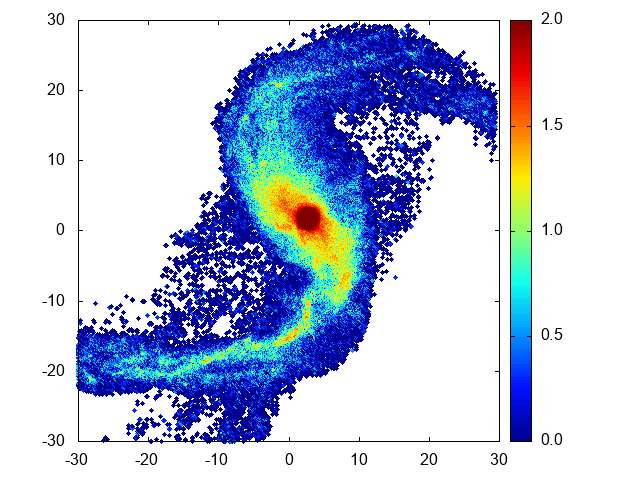}
\includegraphics[width = 1.42in]{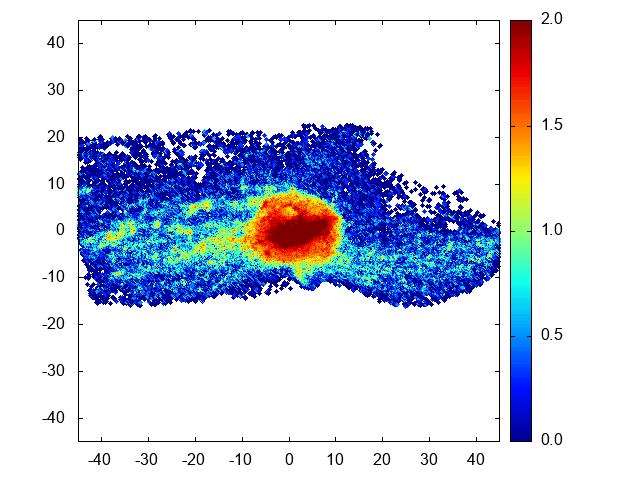}
\includegraphics[width = 1.42in]{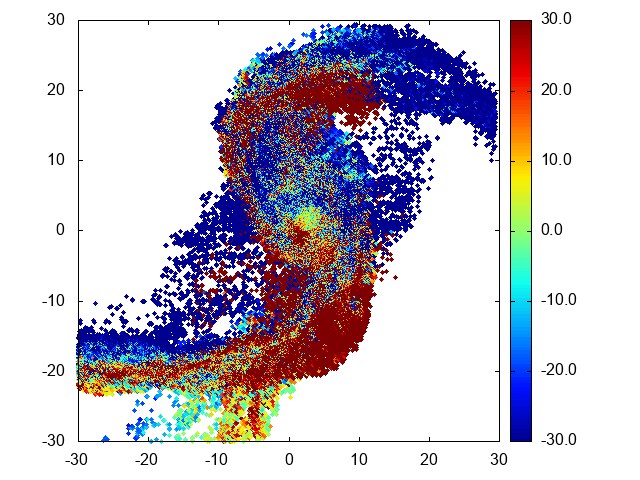}
\includegraphics[width = 1.42in]{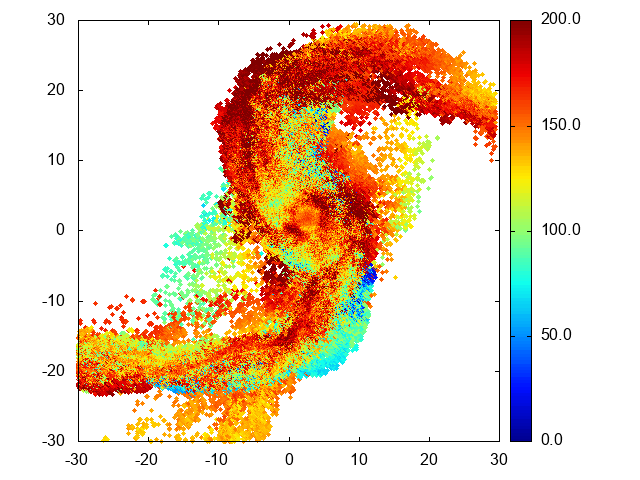}
\includegraphics[width = 1.42in]{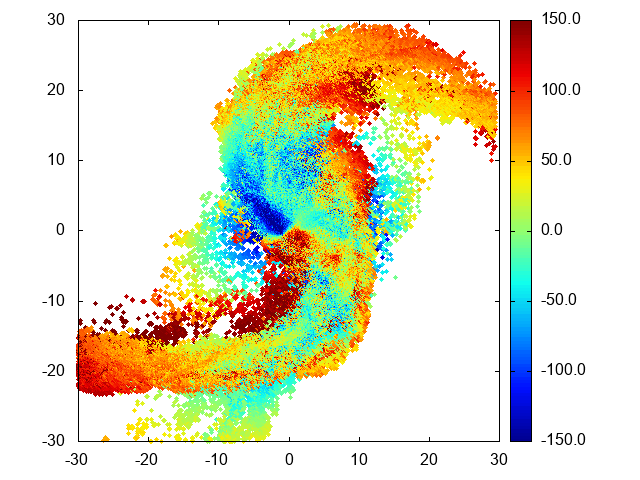}\\
\caption{Upper five panels Run 1, bottom five panels Runs 2 (see text for
  details). From left to right: First panel: projection on the $X-Y$ plane (the color
  code corresponds to the log of the density).  Second panel: shows
  projection on the $X-Z$ plane (the color code corresponds to the log
  of the density).  Third panel: vertical velocity on the $X-Y$ plane
  (the color code corresponds to the velocities in km/s); Fourth
  panel: azimuthal velocity on the $X-Y$ plane (the color code
  corresponds to the velocities in km/s).  Fifth panel: radial
  velocity on the $X-Y$ plane (the color code corresponds to the
  velocities in km/s). Distances are in kpc. The time is $t=0.6$
  Gyr.}
\label{Fig-A2-col3} 
\end{figure*}

Systems formed from the monolithic 
collapse of an isolated over-density both with and without gas dynamics,
show velocity gradients that are generated
during the collapse itself, and evolve in time in a nontrivial way.
As mentioned above, as the IC break spherical symmetry, the collapse
time is different in different directions. In this way, the mechanism
of energy gain is also dependent of the direction. For this reason, the
system resulting from such a collapse not only shows the
heterogeneous velocity field outlined above, but it is
characterized by large scale streaming motions.  A detailed analysis
of this class of systems will be presented in a forthcoming work
\citep{FSL_2019} 
 but we show in Fig.\ref{Fig-A2-col3} two examples. The
IC consists in a perfect oblate ellipsoid (Run 1) and in
an overdensity with an irregular shape (Run 2), respectively.
The mass of the systems is $M= 1.3 \cdot 10^{10} M_\odot$, their
initial gravitational radius is $R_g \approx 10$ kpc, and they have the same amount of angular momentum.  In this way, their collapse
time is $\tau \approx 0.1 \;\; \mbox{Gyr}$. It is important to note that, while this
value of the collapse time is fixed by the choice of $M$ and $R_g$,
and thus may be tuned by changing these parameters, the order of the
magnitude of the velocity components depend on different properties of
the overdensities, most notably on their shape.

Fig.\ref{Fig-A2-col3} shows the two runs at $t=0.6$ Gyr (i.e., $\approx
6 \; \tau$): the projection on the $X-Y$ plane (the systems rotates
around $Z$) shows the dense central region (where motion is close to
isotropic) surrounded by a less dense flat region where the motion is
dominated by the tangential velocity and finally a more sparer
outermost region where the radial velocity is dominant. In both cases,
there are large scale gradients in the tangential and in the radial
velocity but clearly in the case of Run 2, because the IC was less
axisymmetric, they have larger amplitude. In particular the radial
velocity presents a change from negative to positive, which is an
imprint of the collapsing phase where some particles have gained
kinetic energy and thus positive radial velocities, and other
particles still have radial velocities oriented toward the center and
are thus negative. 
In addition, the vertical velocity also shows large
scale gradients, although of smaller amplitude. The time scale of
surviving the transient structures depends on the amplitude of the
velocity gradients. For instance, a radial velocity gradient of
$\Delta v_R \sim 40$ km/s implies a change in the disk structure on a
time scale on the order of 1 Gyr, given that $\Delta v_R$ is on the
order of 40 kpc/Gyr and the disk radius is $\sim 40$ kpc; larger
velocity gradients imply a faster evolution.

%

\section{Discussion and conclusions}

We explore a variety of dynamical factors that can be tested through the comparison with the extended kinematic maps of Gaia-DR2.
Radial and vertical velocities are analyzed here;  the analysis of the azimuthal velocities and the derivation of the rotation speed will be treated in a different paper (Paper III). 

Many mechanisms may generate either non-null radial velocities or non-null vertical velocities, or there may even be 
a common origin of some vertical and radial waves \citep{Fri19}. The gravitational influence of components of the Galaxy that are different from the disk, such as the long bar or bulge, spiral arms, or tidal interaction with Saggitarius dwarf galaxy, may explain some features of the velocity maps, especially in the inner parts of the disk. However, they are not sufficient in explaining the most conspicuous gradients in the outer disk.
Vertical motions might be dominated by external perturbations or mergers, although with a minor component due to the warp, as also concluded in the analysis by \citet{wang2019a,wang2019b} with LAMOST+Gaia data. We carried out N body simulations to investigate the possible contribution of the minor merger to the vertical asymmetrical bulk motions, and we find that the minor merger with nine satellites can explain the positive vertical velocity on both the north and south side of the hemisphere in the range of 8-18 \,kpc.

There are two contributions of the warp to the vertical motion: one was produced by the inclination of the orbits, another contribution was from the variation of the amplitude of the warp angle $\gamma $. Here, we have detected a non-null variation of the relative amplitude of the warp
($\dot{\gamma }/\gamma $) that is significant at the 2.6$\sigma $ level, which is the most likely period for the oscillation of the warp around 0.5 Gyr, although much longer periods are not excluded. Transient warps may be related to a variable torque over the disk; for instance, when the torque is produced by the misalignment of the halo and disk and when the realignment is produced in less than 1 Gyr \citep{Jia99} or in a scenario of accretion of intergalactic matter with variable  ratios of accretion \citep{Lop02b}.

Kinematics distributions, including information on the dispersion of velocities, can also be related with the width of the disk. Here,
we see with two different methods that the mass distribution of the disk is flared, that is, the thickness of the disk increases outwards.
The obtained numbers are roughly consistent with previous analysis of the flare based only on the morphology \citep[and references therein]{martin_flare}, so we can connect both increasing scaleheight and dispersion of velocities outwards as the same phenomenon.
Nonetheless, we must also consider that nonequilibrium systems do not strictly follow the Jeans equation that we have
applied in previous sections. \citet{Hai19} show that traditional Jeans modeling should give reliable results in overdense regions of the disk, but important biases in underdense regions call for the development of nonequilibrium methods to estimate the dynamical matter density locally and in the outer disk. Further analyses of this deviation of Jeans equation in nonequilibrium systems for the application of the present Gaia data will be explored in Paper III.

Precisely, the nonequilibrium system is one of the conclusions of our work here. In lack of other possible causes for the main observed features in the kinematical maps, they can only find an explanation in terms of models in which the Galactic disk is still in evolution, either because the disk has not reached equilibrium since its creation or because external forces, such as the Sagittarius dwarf galaxy, might perturb it.
Here, we explore a simple class of out-of-equilibrium, rotating, and asymmetrical mass distributions that evolve under their own gravity. Noncircular orbits and with significant vertical velocities in the outer disk are precisely one of the predictions of this model. Orbits in the very outer disk are out of equilibrium so they have not reached circularity yet, precisely as we observe in our data. 
The large velocity gradients observed in the Gaia DR2 data are at odds with the simple model in which stars move on steady circular orbits around the center of the Galaxy. The nonequilibrum model that we discuss provides a first and qualitative framework in which such nonstationarity is intrinsic to the dynamical history of the Galaxy rather than being induced by an ad hoc external field due to the passage of a satellite galaxy.

Certainly, further kinematic information at farther distances or along different lines of sight might better constrain the dynamical scenarios of our Galaxy. Future data releases of Gaia will improve our measurements and analyses, especially if we apply the techniques
of extension of kinematic maps using the Lucy's method for the deconvolution of the parallax errors, as was done in Paper I.
On the one hand, the  Gaia mission DR3 will provide with more accurate measurements of the parallaxes so that we can make a direct test on the Lucy's method used on the DR2.  On the other hand, such data will allow us to explore the outermost part of the disk for $R>20$ kpc where larger velocity gradients are expected according to the nonequilibrium models we have discussed in this work.
Also in other galaxies, two-dimensional spectroscopy (for instance, or radio data of THINGS \citep{Syl19a}, or using Multi Unit Spectroscopic Explorer at the Very Large Telescope may allow us to carry out an analysis of noncircularity in mean orbits. 
However, for the case of external galaxies, there is an intrinsic degeneracy between radial and rotational motions if axisymmetry is broken and thus only the direct measurements of the 3D velocity field can clarify the nature of these systems.


\begin{acknowledgements}
Thanks are given to S. Comer\'on for helpful comments, and Rachel Rudy (language editor of A\&A) for proofreading of this text.
MLC, FG and ZC were supported by the grant PGC-2018-102249-B-100 of the Spanish Ministry of Economy and Competitiveness (MINECO).
HFW is supported by the LAMOST Fellow project and funded by
China Postdoctoral Science Foundation via grant 2019M653504.
FSL acknowledges the financial support of the project DynSysMath of the Istituto Italiano di Fisica Nucleare (INFN); this work was granted access to the HPC resources of The
Institute for Scientific Computing and Simulation financed by
Region Ile de France and the project Equip@Meso (Reference
No. ANR-10-EQPX-29-01) overseen by the French National
Research Agency as part of the Investissements d'Avenir program.
RN was supported by the Scientific Grant Agency VEGA No. 1/0911/17. 
This work has made use of data from the European Space Agency (ESA) mission
{\it Gaia} (\url{https://www.cosmos.esa.int/gaia}), processed by the {\it Gaia}
Data Processing and Analysis Consortium (DPAC,
\url{https://www.cosmos.esa.int/web/gaia/dpac/consortium}). Funding for the DPAC
has been provided by national institutions, in particular the institutions
participating in the {\it Gaia} Multilateral Agreement.
\end{acknowledgements}

\end{document}